\documentclass[sigconf]{acmart}

\usepackage{mathrsfs,subfigure}
\usepackage{diagbox}
\usepackage{enumitem}
\usepackage{multirow}
\usepackage{booktabs}
\usepackage{adjustbox}
\usepackage{colortbl}
\newcommand{\tabincell}[2]{\begin{tabular}{@{}#1@{}}#2\end{tabular}}  

\AtBeginDocument{%
  \providecommand\BibTeX{{%
    \normalfont B\kern-0.5em{\scshape i\kern-0.25em b}\kern-0.8em\TeX}}}

\setcopyright{none}
\renewcommand\footnotetextcopyrightpermission[1]{}

\begin{document}



\title{An Empirical Study on Serverless Workflow Services}



\newcommand{\para}[1]{\smallskip\noindent{\bf {#1}. }}
\newcommand{\czp}[1]{{\color{red}{#1}}}
\newcommand{\xzl}[1]{{\color{blue}{#1}}}
\newcommand{\ly}[1]{{\color{purple}{#1}}}
\acmConference[]{}{}{}

\author{Jinfeng Wen}
\affiliation{%
  \institution{Key Lab of High-Confidence Software Technology, MoE (Peking University)}
  \country{Beijing, China}
}
\email{jinfeng.wen@stu.pku.edu.cn}

\author{Yi Liu}
\affiliation{%
  \institution{Key Lab of High-Confidence Software Technology, MoE (Peking University)}
  \country{Beijing, China}
}
\email{liuyi14@pku.edu.cn}
\begin{abstract}
Along with the wide-adoption of Serverless Computing, more and more applications are developed and deployed on cloud platforms. Major cloud providers present their serverless workflow services to orchestrate serverless functions, making it possible to perform complex applications effectively. A comprehensive instruction is necessary to help developers understand the pros and cons, and make better choices among these serverless workflow services. However, the characteristics of these serverless workflow services have not been systematically analyzed. To fill the knowledge gap, we survey four mainstream serverless workflow services, investigating their characteristics and performance. Specifically, we review their official documents and compare them in terms of seven dimensions including programming model, state management, etc. Then, we compare the performance (i.e., execution time of functions, execution time of workflows, orchestration overhead of workflows) under various experimental settings considering activity complexity and data-flow complexity of workflows, as well as function complexity of serverless functions. Finally, we discuss and verify the service effectiveness for two actual workloads. Our findings could help application developers and serverless providers to improve the development efficiency and user experience.
\end{abstract}



\keywords{serverless workflow; serverless computing; empirical study}

\maketitle

\section{Introduction}\label{intro}

Serverless computing is a new paradigm in cloud computing that allows developers to develop and deploy applications on cloud platforms without having to manage any underlying infrastructure, e.g., load-balancing, auto-scaling, and operational monitoring~\cite{carreira2018case, de2016hierarchical, Malawski2020, jonas2017occupy, chard2017ripple, bila2017leveraging, fouladi2017encoding}. Due to its significant advantages, serverless computing has been an increasingly hot topic in both academia~\cite{akkus2018sand, mcgrath2017serverless, jonas2019cloud} and industry~\cite{shahradserverless2020, Amazon, Microsoft, Google}; its market growth is expected to exceed \$8 billion per year by 2021~\cite{FaaSMarket}. In serverless computing, developers prototype an event-driven application as a set of interdependent functions (named as \textit{serverless functions}), each of which performs a single logical task~\cite{datta2020valve}. 
To facilitate the coordination among these functions, in recent years, major cloud providers have rolled out serverless workflow services (e.g., AWS Step Functions~\cite{ASF}), which aim to orchestrate serverless functions in a reliable way. Specifically, serverless workflow allows developers to only specify the execution logic among functions, without having to implement this logic through complex nested function calls~\cite{akkus2018sand}.

With the help of serverless workflow, complex application scenarios (e.g., data processing pipeline and machine learning pipeline) can be accomplished more efficiently~\cite{SWbestpractices}. Given the surging interest in serverless computing and the increasing dependence of current serverless computing on serverless workflow, characterizing existing serverless workflow services is of great significance. On one hand, it can help developers understand the pros and cons of these services and thus make better choices among them according to the application scenarios. On the other hand, it can provide insightful implications for cloud providers to improve these services in a more targeted manner. However, to the best of our knowledge, the characteristics of these serverless workflow services have not been systematically analyzed.

To fill the knowledge gap, this paper presents the first empirical study on characterizing and comparing existing serverless workflow services. Specifically, we focus our analysis on four mainstream serverless workflow services, including AWS Step Functions (ASF), Azure Durable Functions (ADF)~\cite{ADF}, Alibaba Serverless Workflow (ASW)~\cite{SW}, Google Cloud Composer (GCC)~\cite{GCC}. We first review their official documents and compare them in terms of six dimensions including orchestration way, data payload limit, parallelism support, etc. Then, we compare the performance (including execution time of functions, execution time of workflows, orchestration overhead time of workflows, etc.) of the four services under varied experimental settings. The comparison is performed in two representative application scenarios, i.e., sequence applications and parallel applications, which refer to applications that can be prototyped as multiple functions executing in a sequence and parallel way, respectively. Sequence applications and parallel applications can be represented as sequence workflows and parallel workflows, respectively. More specifically, we focus on the following three aspects that we can provide insights for developers and cloud providers:

\begin{itemize}[leftmargin=*]
\item \textbf{The effect of activity complexity.} We first compare the performance of the selected serverless workflow services under various levels of activity complexity~\cite{cardoso2006approaches} (i.e., the numbers of serverless functions contained in a workflow).  
We find that the execution time of workflows, execution time of functions, and orchestration overhead time of workflows all become longer for ASF, ADF, ASW, and GCC with the increase of activity complexity in both sequence and parallel workflows, except that the orchestration overhead time of workflows of GCC has certain fluctuation in parallel workflows.  Additionally, in sequence workflows, we find that the execution time of workflows in ASF, ADF, and ASW are mainly generated by the execution time of functions, whereas GCC is the orchestration overhead of workflows. In parallel workflows, when more functions are required, the execution time of functions gradually increases, and it determines the changing trend of the execution time of workflows. 
\item \textbf{The effect of data-flow complexity.} We then compare the performance of four serverless workflow services under different levels of data-flow complexity~\cite{cardoso2006approaches} (i.e., the size of data payloads passed among functions). We find that only under high data-flow complexity conditions will ASF, ADF, and ASW have a certain impact on the performance in sequence and parallel workflows, while GCC is affected by whether there is a payload or not. \item \textbf{The effect of function complexity.} We also compare the performance of these serverless workflow services under different levels of function complexity (i.e., the specified duration time of serverless functions). We find that the execution time of workflows and execution time of functions become longer for ASF, ADF, and ASW as function complexity increases in sequence and parallel workflows, whereas there is no obvious impact on GCC. Besides, we find that the orchestration overhead of workflows is less affected by function complexity.
\end{itemize}

Based on the derived findings, we have drawn insightful implications for developers and cloud providers. Specific findings and implications are shown in Table~\ref{tab:findingandimplication}. We also offer the source code\footnote{They will be made public later.} used in this study as an additional contribution to the research community for other researchers to replicate and build upon.

\begin{table*}[htb]
 \centering
 \small
 \caption{Summary of findings and implications}
 \begin{adjustbox}{width=2.1\columnwidth}
    \begin{tabular}{|p{1.05\columnwidth}|p{1.05\columnwidth}|}
\hline

\rowcolor{gray!50}
\multicolumn{2}{|c|}{\textbf{Activity complexity} }  \\
\hline

\rowcolor{gray!15}
\textbf{Findings} & \textbf{Implications} \\
\hline

\textbf{F.1}: The total execution time of workflows, execution time of functions, and orchestration overhead time of workflows become longer for ASF, ADF, ASW, and GCC with the increase of activity complexity in both sequence and parallel workflows, except that the orchestration overhead time of workflows of GCC has certain fluctuation in parallel workflows.
& 
\textbf{I.1}: (i) For \textbf{developers}, we advise to select the serverless workflow service with the best performance whether in both sequence and parallel workflows. (ii) For \textbf{developers}, considering the execution time of workflows, we advise ADF is used in small-scale ($\leq$ 10) parallel workflow tasks, ASW is used in large scale (between 10 and 100) tasks, and ADF is used to larger-scale ( > 100) tasks.\\ \hline

\textbf{F.2}: In sequence workflows, the execution time of workflows in ASF, ADF, and ASW are mainly generated by the execution time of functions, whereas GCC is the orchestration overhead time of workflows.
&
\textbf{I.2}: The \textbf{cloud provider} of GCC should analyze the framework of workflow scheduling, and improve significantly the efficiency of the workflow execution of Google Cloud Composer.
\\ \hline

\textbf{F.3}: When more functions participate into sequence workflows, the orchestration overhead of workflows gradually increases, and the orchestration overhead determines the changing trend of the total time of sequence workflows.
& 
\textbf{I.3}: For \textbf{developers}, considering the execution time and orchestration overhead time of workflows, we advise ASF is used in activity-intensive sequence workflow tasks.
\\ \hline
\textbf{F.4}: When more functions participate into parallel workflows, the execution time of functions gradually increases, and the execution time of functions determines the changing trend of the total time of parallel workflows. 
& 
 \textbf{I.4}: \textbf{Cloud providers} of different serverless workflow services should further improve the efficiency of the scheduling algorithms among functions to reduce the function execution overhead in parallel workflows.
\\ \hline

\hline

\rowcolor{gray!50}
\multicolumn{2}{|c|}{\textbf{Data-flow complexity} }  \\
\hline

\rowcolor{gray!15}
\textbf{Findings} & \textbf{Implications} \\
\hline

\textbf{F.5}: Regarding to the data payload limit, different serverless workflow services are different and do not have uniform limit, i.e., ASF is $2^{18}$B, ASW is $2^{15}$B, GCC has internal storage limit with $2^{15}$B, and ADF does not have this restriction. 
& 
 \textbf{I.5}: For \textbf{cloud providers} of different serverless workflow services, they should exchange technologies with each other to break through their respective implementation bottlenecks of the data payload.
\\ \hline

\textbf{F.6}: ASF, ADF, ASW have a little impact under low data payload conditions. Only under high data payload conditions will ASF, ADF, ASW have a certain impact, whereas GCC is affected by whether there is a data payload or not.
&
\textbf{I.6}: For \textbf{developers}, considering the execution time and orchestration overhead time of workflows, we advise ASF is used in data-flow-intensive sequence (or parallel) workflow tasks, where payloads are less than $2^{18}$B (or $2^{15}$B).\\

\hline

\textbf{F.7}: ADF can pass a larger data payload (e.g., $2^{20}$B). If developers continue use ASF, they can store the data payloads in Amazon S3, then use the Amazon Resource Name (ARN) instead of raw data in serverless workflows. 
& 
 \textbf{I.7}: For \textbf{developers}, when a workflow task needs to pass a larger data payload (> $2^{18}$B in sequence workflows, or > $2^{15}$B in parallel workflows ), we advise to use ADF, or ASF with the external storage.
\\ \hline

\rowcolor{gray!50}
\multicolumn{2}{|c|}{\textbf{Function complexity} }  \\
\hline

\rowcolor{gray!15}
\textbf{Findings} & \textbf{Implications} \\
\hline

\textbf{F.8}: The function execution of ADF performs better than other serverless workflow services whether in sequence or parallel workflows.
&
 \textbf{I.8}: \textbf{Cloud providers} of ASF, ASW, and GCC need to improve the performance of serverless functions on their serverless computing platforms.\\ 
 \hline
 
\textbf{F.9}: The orchestration overhead of workflows is less affected by changes within serverless functions (function complexity), but more affected by changes in the workflow structure (activity complexity) or data payload (data-flow complexity).
& 
 \textbf{I.9}: For \textbf{developers}, considering the results of the execution time and orchestration overhead time of workflows, we advise ASF (or ADF) is used in function-sensitive sequence (parallel) workflow tasks.
 \\ \hline

\hline

\end{tabular}
\end{adjustbox}
\label{tab:findingandimplication}
\end{table*}

\section{Feature Comparison of serverless workflow Services}\label{characteristic}

We first select four mainstream serverless workflow services from public cloud platforms, including AWS Step Functions (ASF) (released December 1, 2016), Azure Durable Functions (ADF) (released May 7, 2018), Alibaba Serverless Workflow (ASW) (released July 2019), Google Cloud Composer (GCC) (released May 1, 2018). These services have relatively mature application practices and are more standardized rather than those of private cloud platforms. Then, through reviewing official documents, we compare the features of these serverless workflow services from the following dimensions: 
\begin{itemize}[leftmargin=*]
\item {Orchestration way}: the workflow definition model and model definition language of serverless workflow services.
\item {Data payload limit}: the size constraint of data payloads transmitted among serverless functions of a serverless workflow.
\item {Parallelism support}: whether serverless workflow services support to parallel multiple serverless functions.
\item {Execution time limit}: the maximum execution time of workflows supported by serverless workflow services.
\item {Reusabiluty}: whether a serverless workflow can be used to a part (called sub-workflow) of another serverless workflow.
\item {Supported development language}: the supported development languages for serverless workflow services.
\end{itemize}
Table~\ref{tab:comparefeature} shows the results of the feature comparison as of Sep. 2020. In detail, the feature of each dimension is explained as follows:

\begin{table*}[htb]
 \centering
 \small
 \caption{A feature comparison in four serverless workflow services.}
 \begin{adjustbox}{width=2.1\columnwidth}
    \begin{tabular}{|p{0.3\columnwidth}|p{0.4\columnwidth}|p{0.4\columnwidth}|p{0.5\columnwidth}|p{0.4\columnwidth}|}
\hline

\rowcolor{gray!50}
& \textbf{AWS Step Functions}
 & \textbf{Azure Durable Functions}
 & \textbf{Ablibaba Serverless Workflow}
 & \textbf{Google Cloud Composer}  \\
\hline

\cellcolor{gray!15}\textbf{Orchestration way}	& State Machine / State Definition Language (JSON) & Orchestrator Function / In code &  Flow / Flow Definition Language (JSON) &  Directed Acyclic Graph / In code \\
\hline

\cellcolor{gray!15}\textbf{Data payload limit}	& 256KB & Unknown & 32KB & Unknown  \\
\hline

\cellcolor{gray!15}\textbf{Parallelism support}	& Yes & Yes  & Yes  & Unknown \\
\hline

\cellcolor{gray!15}\textbf{Execution time limit}	& \tabincell{l}{Standard type: 1 year \\ Express: 5 minutes} & Unlimited & 1 year & Unlimited \\
\hline

\cellcolor{gray!15}\textbf{Reusability}	& Yes & Yes & Yes & No \\
\hline

\cellcolor{gray!15}\textbf{Supported development language}	& Java, .NET, Ruby, PHP, Python (Boto 3), JavaScript, Go, C++ & C$\#$, JavaScript, F$\#$, PowerShell, Python & Java, Python, PHP, .NET, Go, JavaScript &  Python 2, Python 3 \\
\hline
\end{tabular}
\end{adjustbox}
\label{tab:comparefeature}
\end{table*}

\noindent\textbf{Orchestration way.} We explain the orchestration way of serverless workflow services in terms of two perspectives. From the perspective of the workflow definition model, each serverless workflow service uses the respective workflow model. Specifically, ASF and ASW are based on \emph{State Machine}\footnote{A state machine is just a collection of states, the relationship among the states and their inputs and outputs.} and \emph{Flow}\footnote{Flow defines the business logic description and the general information required to execute the process.}, respectively, whereas ADF uses a new type of function (called \emph{Orchestrator Functions}\footnote{Orchestrator functions are the heart of ADF and describe the order in which actions are executed.}) and GCC defines workflows by creating a Directed Acyclic Graph (DAG\footnote{A DAG is a collection of tasks organized to reflect their directional interdependencies.}). From the perspective of model definition language, developers write workflow models of ASF and ASW through the JSON format, whereas ADF and GCC need to leverage the procedural code. Specifically, the definition languages of the workflow model in ASF and ASW are \emph{Amazon State Language}\footnote{\url{https://states-language.net/spec.html}} and \emph{Flow Definition Language}\footnote{\url{https://help.aliyun.com/document_detail/122492.html?spm=a2c4g.11186623.6.575.fecf52c2lWEbbq}, in Chinese}, respectively. \emph{State Definition Language} and \emph{Flow Definition Language} are both the JSON format. However, in ADF, developers design and orchestrate serverless functions in the code of the orchestrator function, whereas DAG of GCC is written in a \emph{Python} script.

\noindent\textbf{Data payload limit.} Official documents of ASF and ASW specifically explain the data payload constraints in workflows, whereas ADF and GCC do not. Specifically, in ASF, the maximum input or result data size for a task or execution defaults to 256KB~\cite{ASFpayload}. Differently, for ASW, the total size of the input, output, and local variables cannot exceed 32KB. Note that the local variable is used to store the output of functions, and we have verified that the data size of the local variable will account for part of the total data size in our experiments. Although documents of ADF and GCC do not mention their data constraints, we find some interesting phenomenons in our later experiments. For example, ADF can achieve a larger data transmission (i.e., 1024KB) than ASF (i.e., 256KB). For GCC, without the help of external storage, native data will be limited by the 32KB of its database storage built in the Cloud Composer environment.

\noindent\textbf{Parallelism support.} ASF, ADF, and ASW support to parallel serverless functions. However, regarding the parallel structure in GCC, there is no explicit mention in its document. Though reviewing the operators relationship of the DAG concept~\cite{DAGrelationship}, we find that the \emph{list} relationship can express the function parallelism.

\noindent\textbf{Execution time limit.} ASF and ASW have the execution time limitation of workflows, whereas ADF and GCC are not limited and they can execute for a long time. Specifically, when creating a state machine in ASF, two types~\cite{StandardvsExpressWorkflows} can choose: (i) \emph{Standard} type that can run for up to one year and is ideal for long-time, durable, and auditable workflows; (ii) \emph{Express} type that can run for up to five minutes and is ideal for high-volume, event-processing workloads. Additionally, ASW supports the flow execution for up to one year.

\noindent\textbf{Reusability.} ASF, ADF, and SW can integrate with their own API to form nested workflows. When building new workflows, using nested workflows can reduce the complexity of the main workflow. However, GCC cannot integrate and reuse its workflows because it relies on a managed \emph{Airflow} deployment. The \emph{Airflow} is an approach for workflow management on a dedicated long-running workflow-execution engine. Its existence indicates that the DAGs composed of multiple tasks cannot be treated as serverless functions.

\noindent\textbf{Supported development language.} The supported development language of each serverless workflow service is different. Specifically, (i) ASF is supported by the AWS SDKs for Java, .NET, Ruby, PHP, Python (Boto 3), JavaScript, Go, and C++. (ii) ADF currently supports C$\#$, JavaScript, F$\#$, PowerShell, and Python. (iii) ASW provides SDKs for Java, Python, PHP, .NET, Go, and JavaScript. Furthermore, (iv) GCC only supports Python 2 and Python 3.

\section{Methodology}\label{method}

\begin{figure}[h]
  \centering
  \includegraphics[width=0.45\textwidth]{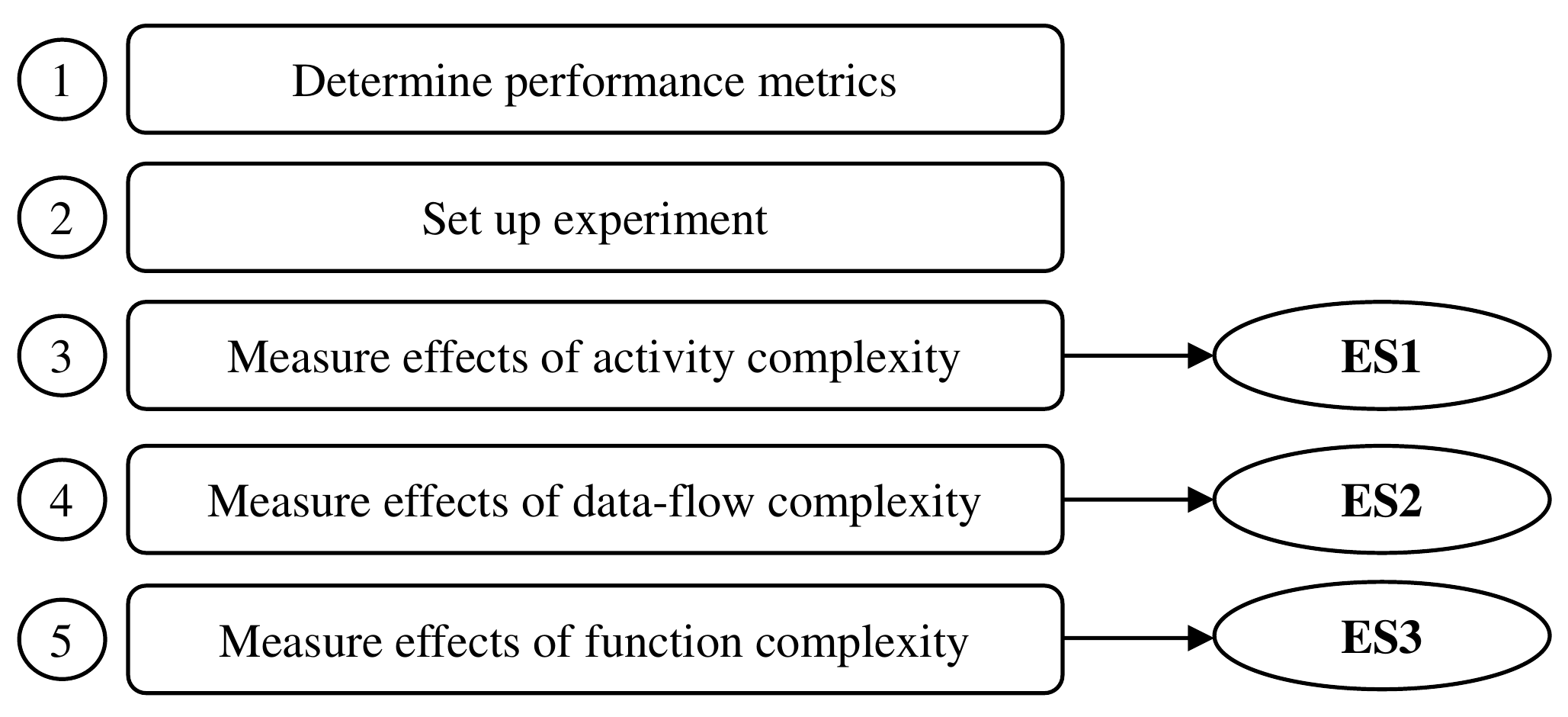}
  \caption{An overview of the methodology.}
  \label{fig:overview}
\end{figure}

We consider two representative scenarios, i.e., sequence applications and parallel applications, which refer to applications that can be prototyped as multiple functions executing in a sequence and parallel way, respectively. Sequence applications and parallel applications can be represented as sequence workflows and parallel workflows, respectively. To measure and compare the performance of serverless workflow services under varied experimental settings, we show an overview of the methodology of our study in Figure~\ref{fig:overview}.

\noindent\emph{\textbf{Step 1: Determine performance metrics.}} In the first step, we determine the performance metrics of our study. Generally, time for the process of workflow executions is spent on the execution of workflows and functions, as well as orchestration overhead of functions. Thus, the metrics related to them are considered, and the specific representations and meanings are explained as follows. (i) \emph{\textbf{totalTime}} is the total execution time to complete a workflow. (ii) \emph{\textbf{funTime}} is the actual execution time of functions contained in workflows. The calculation strategies of \emph{funTime} are different in different application scenarios. Specifically, in sequence workflows, \emph{funTime} is the sum of actual execution times of all functions contained in workflows, whereas \emph{funTime} is the time interval between the start time of the first function execution and the end time of the last function completion in parallel workflows. (iii) \emph{\textbf{overheadTime}} is the actual overhead time produced in the orchestration process of a workflow. \emph{overheadTime} may contain duration times of workflow start, function scheduling, data state transition, parallel branch and merge, etc. (iv) \emph{\textbf{theo$\_$overheadTime}} is the theoretical overhead time produced in the orchestration process of a parallel workflow. Theoretically, paralleling multiple functions costs the time of the function with the longest execution time. The theoretical parallel workflow is a zero-overhead parallel composition. Removing the theoretical specified duration time of functions from the total time of workflows is the theoretical overhead time of workflows. Though comparing \emph{overheadTime} with \emph{theo$\_$overheadTime}, we believe some interesting findings can be found in our study. Note that the sum of \emph{funTime} and \emph{overheadTime} equals to \emph{totalTime} for a workflow.

\noindent\emph{\textbf{Step 2: Set up experiment.}} Most of our experiments were done from June 15 - August 20, 2020. In our study, without considering the cold start of spawning the function containers, serverless functions in workflows are in \emph{warm} state to avoid undesired startup latency. A common practice is to reuse launched containers by keeping them \emph{warm} for a period of time. The first call to a serverless function is still \emph{cold}, but subsequent calls to this serverless function can be served by the \emph{warm} container. Thus, each group experiment is repeated several measurements to ensure the correctness of the results. We discord the result of the first measurement and keep the remaining results to evaluate the final performance of serverless workflow services. Additionally, we adopt the median of remaining results to compare \emph{totalTime}, \emph{funTime}, and \emph{overheadTime} of various serverless workflow services.

Generally, serverless functions contained in workflows are executed on cloud platforms in a distributed manner. In fact, each serverless function is like the black box relative to workflows, and its external manifestation is its input parameter, output result, and its execution time. Workflows do not care about the specific implementation in serverless functions. Thus, we leverage the sleep functionality to simulate the function implementation. In order to ensure the function comparability, we try to write serverless functions in a consistent language. According to the serverless community survey, \emph{Node.js} accounts for 62.9$\%$ in languages used for serverless development, and it is the most popular runtime overall \cite{serverlesscommunitysurvey}. In our study, ASF, ADF, and ASW adopt serverless functions with the \emph{JavaScript} language. Because DAGs of GCC are written by \emph{Python} script, tasks in GCC are achieved by the \emph{Python} language. Due to the simplicity of serverless functions, theoretically, the execution time of functions should not be much different. Thus, GCC can be compared with ASF, ADF, and ASW. Additionally, ASF and ASW need to configure the function size in advance, thus we set as 128MB of the memory uniformly. Regarding the region setting of serverless workflow services, ASF, ADF, and GCC are set to US-west uniformly, whereas ASW chooses Shanghai of China because it is only supported in Asia.

\noindent\emph{\textbf{Step 3: Measure effects of activity complexity.}} Activity complexity describes the number of functions a workflow has~\cite{cardoso2006approaches}. Considering both sequence and parallel workflows, we configure various numbers of functions in workflows. Because the maximum number of branches in parallel is limited to 100 in ASW, the function number of our study specifies as 2, 5, 10, 20, 40, 80, 100, and 120. Particularly, the experimental test about the function number with 120 is to verify the parallel limitations of ASF, ADF, and GCC, because their documentations do not explicitly indicate.  Shahrad \textit{et al.}~\cite{shahradserverless2020} presented that the distribution of function execution times on Azure Functions~\cite{AzureFunctions} shows a sufficiently log-normal fit to the distribution of the average function execution time. We find that the probability that the function execution time is one second is the greatest, thus it illustrates that a majority of serverless functions implement the functionality with one-second execution. In experiments of the activity complexity, we set all serverless functions to sleep for one second. The \emph{Step 3} corresponds to the details of the experimental setting \textbf{ES1}.

\noindent\emph{\textbf{Step 4: Measure effects of data-flow complexity.}} Data-flow complexity reflects on the date payload used in effect pre- and post-conditions of function execution~\cite{cardoso2006approaches}. Considering both sequence and parallel workflows, we configure various data payloads of functions for workflows. Table~\ref{tab:comparefeature} shows that the maximum data size between functions in ASW is 32KB ($2^{15}$B). To verify whether ASW can pass a larger data payload, we set data payloads with 0B, $2^5$B, $2^{10}$B, $2^{15}$B, $2^{16}$B. For most of the sequence applications~\cite{SWbestpractices}, we find that about five serverless functions can basically fulfill their requirements unless applications need to add additional functionalities. For example, the video processing needs the following steps: (i) video extraction, (ii) slicing, (iii) transcoding, (iv) merging, (v) post-processing, and additional functionalities (e.g., watermark insertion or information update). Thus, sequence workflows in our study contain five same serverless functions, which each one receives a parameter (i.e., payload), sleeps for one second, and returns this parameter. However, for parallel applications (e.g., data processing pipeline), a large amount of data may need to be processed in parallel. Generally speaking, the number of parallel functions depends on the data scale and expected efficiency. Thus, we cannot determine the function number contained in parallel workflows. Considering that our main purpose is to explore the effects of various payloads on parallel workflows, it is reasonable and comparable as long as different serverless workflow services set the same function number in parallel workflows. In our experiments, we set the five same serverless functions in parallel workflows. The \emph{Step 4} corresponds to the details of the experimental setting \textbf{ES2}.

\noindent\emph{\textbf{Step 5: Measure effects of function complexity.}} Function complexity reflects on the time required to implement a serverless function. Considering both sequence and parallel workflows with five same serverless functions, we configure various specified duration times of serverless functions. Shahrad \textit{et al.}~\cite{shahradserverless2020} mentioned 96$\%$ of serverless functions take less than 60 seconds on average. Thus, we set the specified duration time of sleep functions as 50ms, 100ms, 1s, 10s, 20s, 40s, 60s, and 120s without data payloads.

The \emph{Step 5} corresponds to the details of the experimental setting \textbf{ES3}.

\section{Results}\label{result}

In this section, we show and discuss results under various levels of activity complexity, data-flow complexity, and function complexity considering both sequence and parallel application scenarios. All result values obtained from our experiments are in seconds and figures are available on our Github. Then, we report a series of findings and implications for developers and cloud providers. In our result explanation, we first discuss the execution time of workflows (\emph{totalTime}), and execution times of functions (\emph{funTime}) with the increase of the number of functions, then orchestration overhead time of workflows (\emph{overheadTime}), finally the distributions of measurement results about these three metrics.

\subsection{Activity Complexity (ES1)}

Activity complexity reflects on the numbers of serverless functions contained in a workflow.

\subsubsection{Sequence application scenario}

\begin{figure}[htbp!]
  \centering
  \includegraphics[width=\linewidth]{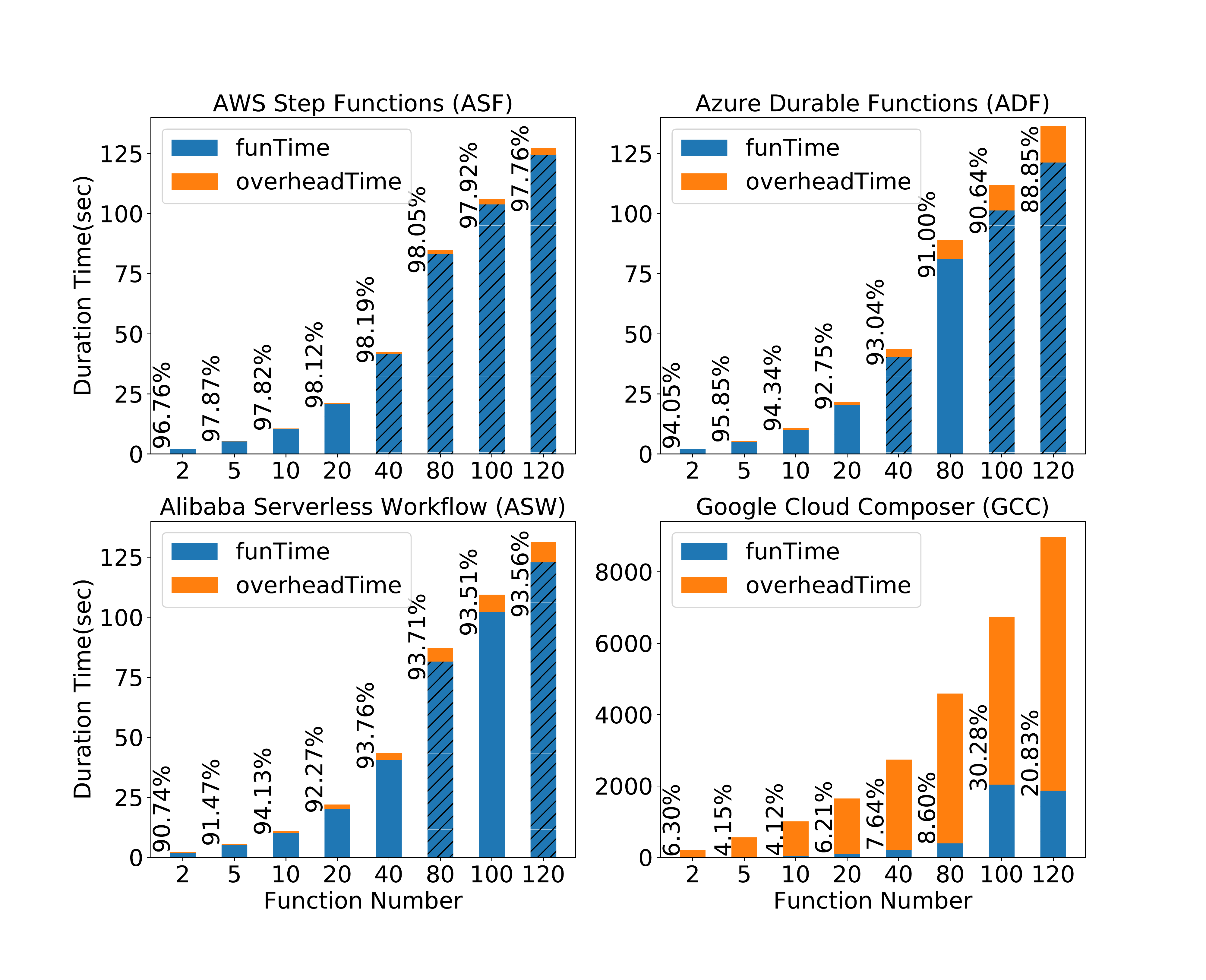}
  \caption{The performance of various levels of activity complexity in sequence workflows.}
  \label{fig:FunctionNumSeq-totalTime+funTime+overheadTime}
\end{figure}

Figure~\ref{fig:FunctionNumSeq-totalTime+funTime+overheadTime} represents \emph{totalTime}, \emph{funTime}, and \emph{overheadTime} about various numbers of functions contained in sequence workflows for ASF, ADF, ASW, and GCC. The horizontal axis is the number of functions, and the vertical axis is the duration time in seconds. Each bar in  Figure~\ref{fig:FunctionNumSeq-totalTime+funTime+overheadTime} consists of \emph{funTime} and \emph{overheadTime} produced from the workflow with a fixed number of functions. Note that the sum of \emph{funTime} and \emph{overheadTime} equals to \emph{totalTime} for this workflow. The value next to the bar indicates the percentage of \emph{funTime} to \emph{totalTime}. 

For \emph{totalTime} and \emph{funTime} in Figure~\ref{fig:FunctionNumSeq-totalTime+funTime+overheadTime}, as more serverless functions are added into sequence workflows, \emph{totalTime} and \emph{funTime} of ASF, ADF, ASW, and GCC both increase. Undoubtedly, when the number of functions contained in sequence workflow increases, \emph{funTime} will inevitably increase, thus \emph{totalTime} increases. Generally speaking, five one-second functions have a \emph{totalTime} of more than five seconds, ten one-second functions are more than ten seconds, etc. We find that \emph{totalTime} of ASF, ADF, and ASW basically conforms to such a growing trend. Besides, \emph{totalTime} of ASF, ADF, and ASW depends on \emph{funTime}. Specifically, the percentage value of ASF fluctuates between 96.76$\%$ and 98.19$\%$, ADF is between 91.00$\%$ and 95.85$\%$, and ASW is between 90.74$\%$ and 94.13$\%$. However, the percentage value of GCC is only between 4.12$\%$ and 8.60$\%$, thus it illustrates that most of the time on GCC is spent on \emph{overheadTime} rather than \emph{funTime}. The main reason may be due to the environment setting itself.

\begin{figure}[ht]
  \centering
  \includegraphics[width=\linewidth]{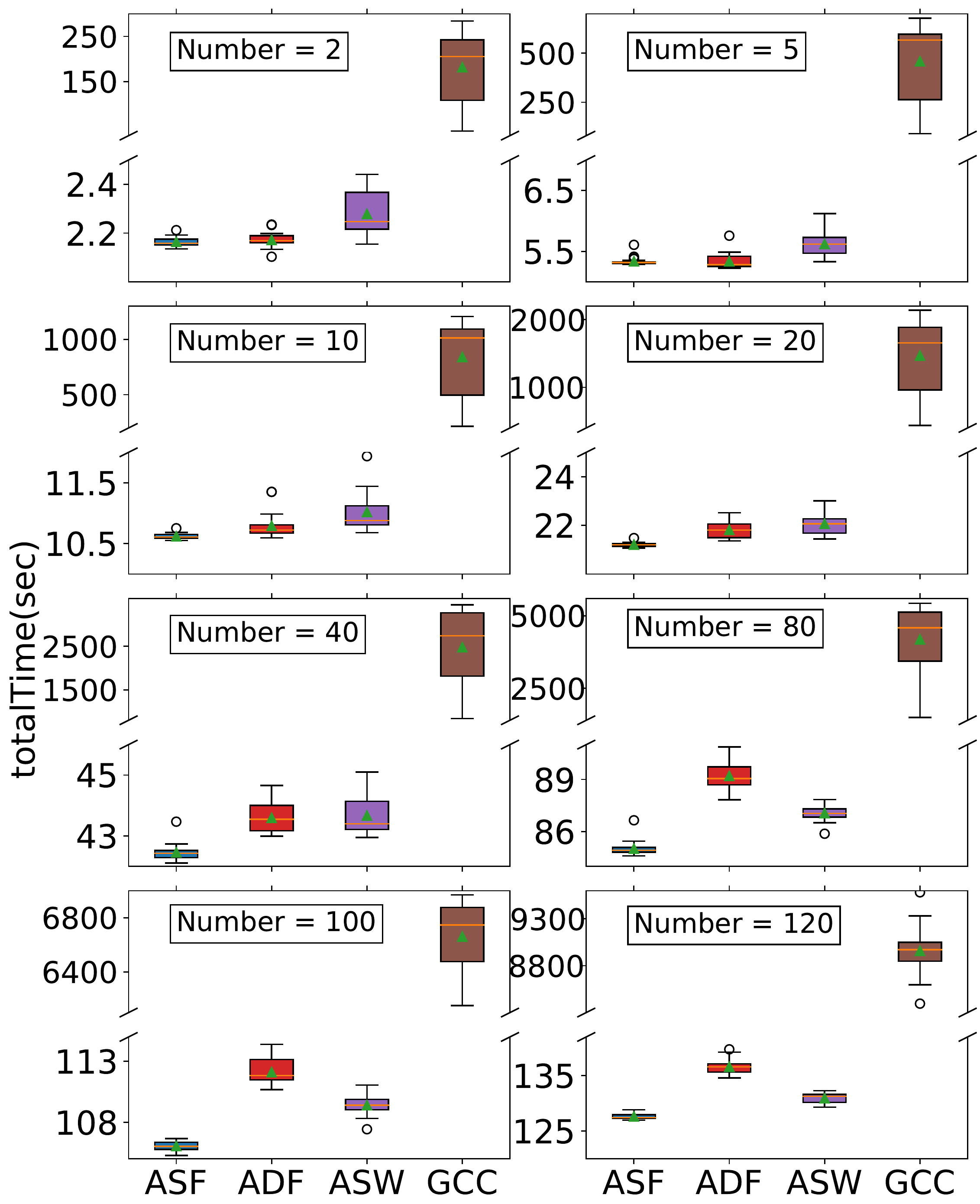}
  \caption{The comparison of \emph{totalTime} under various levels of activity complexity in sequence workflows.}
  \label{fig:FunctionNumSeq-compare-totalTime}
\end{figure}

\begin{figure}[ht]
  \centering
  \includegraphics[width=\linewidth]{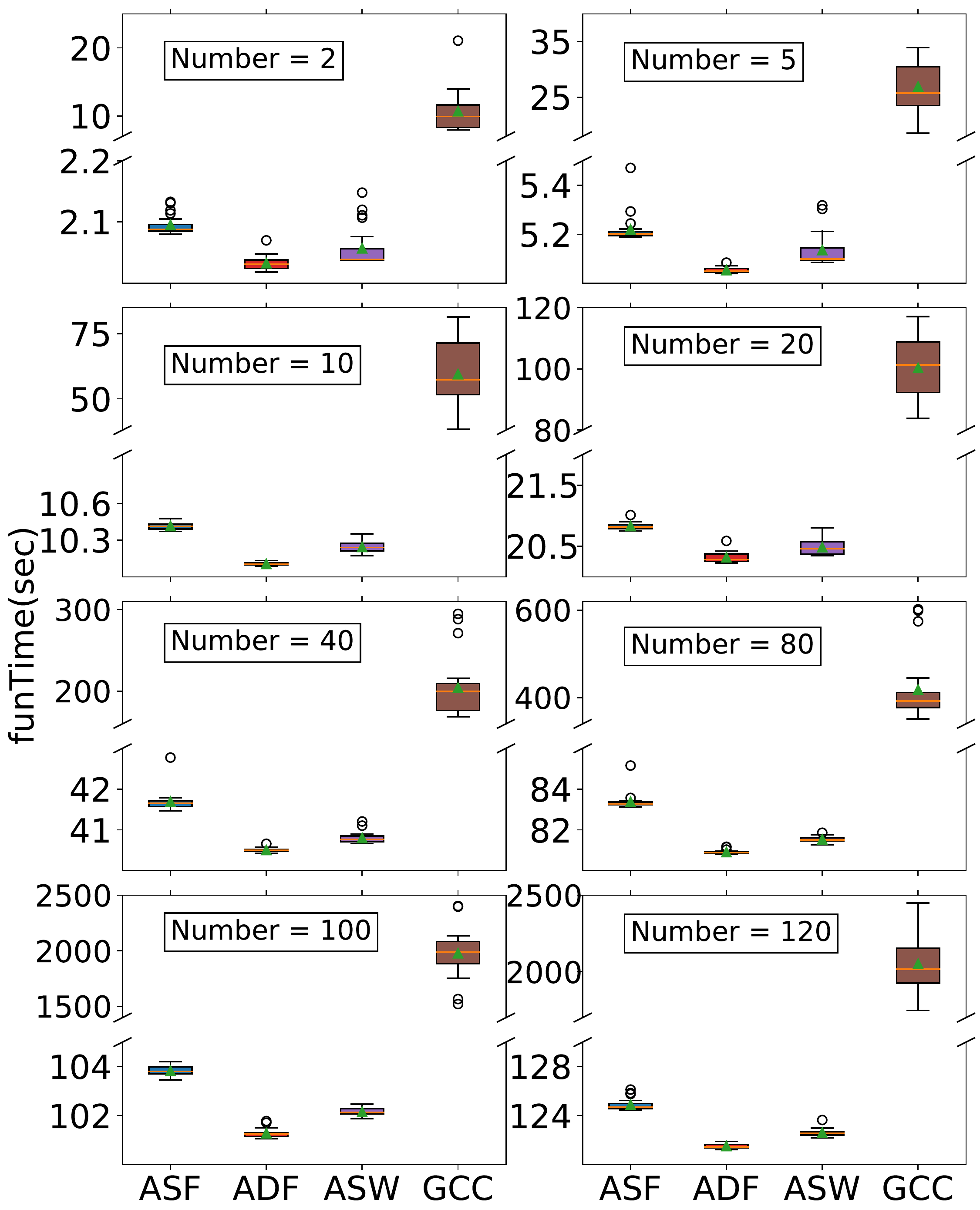}
  \caption{The comparison of \emph{funTime} under various levels of activity complexity in sequence workflows.}
  \label{fig:FunctionNumSeq-compare-funTime}
\end{figure}

For \emph{overheadTime} in Figure~\ref{fig:FunctionNumSeq-totalTime+funTime+overheadTime}, we find that \emph{overheadTime} of ASF, ADF, ASW, and GCC gets longer as more functions are added to sequence workflows. Thus, \emph{the number of functions contained in sequence workflows will affect the orchestration overhead of workflows.} 

To comprehensively compare the performance of ASF, ADF, ASW, and GCC, we display the statistical results of all measurements in a format of the box plot. Figure~\ref{fig:FunctionNumSeq-compare-totalTime} shows that the comparison of \emph{totalTime} under varied numbers of functions for ASF, ADF, ASW, and GCC. We observe that ASF has the lowest and most stable \emph{totalTime}, whereas GCC is the opposite. Additionally, when the number of functions contained in sequence workflows does not exceed 40, the overall result about \emph{totalTime} of ADF is lower than that of ASW. However, when the number of functions increases (larger than 40), \emph{totalTime} of ADF begins to exceed that of ASW. To explore which factors affect \emph{totalTime}, we observe the distribution results of \emph{funTime} and \emph{overheadTime}. We find \emph{funTime} is longer than the theoretical execution time of functions in Figure~\ref{fig:FunctionNumSeq-compare-funTime}, where the origin point of the Y-axis on each sub-graph is the theoretical execution time of functions, i.e., 2s, 5s, 10s, 20s, 40s, 80s, 100s, 120s. Figure~\ref{fig:FunctionNumSeq-compare-funTime} also shows that ADF has the lowest \emph{funTime}, followed by ASW, ASF, and finally GCC. For the \emph{overheadTime} comparison, due to the space reason, its distribution is not displayed. We find that no matter how many functions are in sequence workflows, ASF has often the lowest \emph{overheadTime}, and GCC is still the highest. In particular, for the change trend pf ADF and ASW, it is basically consistent with the comparison of \emph{totalTime} in Figure~\ref{fig:FunctionNumSeq-compare-totalTime}. Thus, \emph{the changing trend of the total time of workflows in sequence workflow is mainly affected by the orchestration overhead of workflows.} Reducing the orchestration overhead is vital for serverless workflow. Strategies about workflow start, state transition, and function scheduling need to be rethought to define by cloud providers.

\subsubsection{Parallel application scenario}

\begin{figure}[htbp!]
  \centering
  \includegraphics[width=\linewidth]{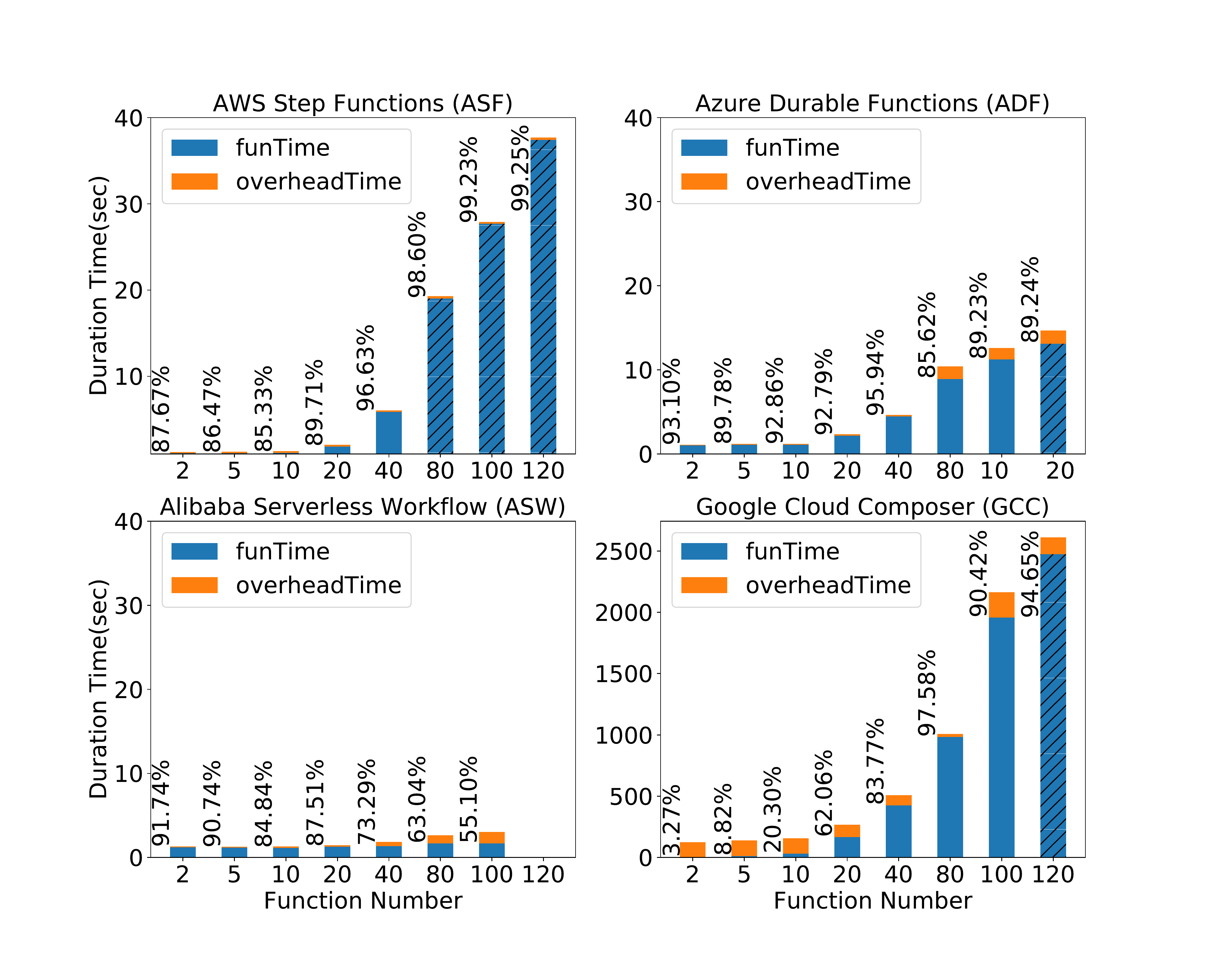}
  \caption{The performance of various levels of activity complexity in parallel workflows.}
  \label{fig:FunctionNumPar-totalTime+funTime+overheadTime}
\end{figure}

Figure~\ref{fig:FunctionNumPar-totalTime+funTime+overheadTime} shows that \emph{totalTime}, \emph{funTime}, and \emph{overheadTime} of varied numbers of functions contained in parallel workflows for ASF, ADF, ASW, and GCC. As more serverless functions are added to parallel workflows, \emph{totalTime} of ASF, ADF, ASW, and GCC is showing an increasing trend. From the percentage values  (the ratio of \emph{funTime} to \emph{totalTime}) in ASF and ADF, we can observe that \emph{totalTime} is mainly used for their \emph{funTime}. Values in ASF ranges from 85.33$\%$ to 99.25$\%$, whereas ADF is 85.62$\%$ to 95.94$\%$. Additionally, for ASW, when the number of functions is small, its \emph{totalTime} is mainly used for function executions. However, when more functions are added into parallel workflows, its proportion values gradually decrease. It illustrates that \emph{overheadTime} is increasing with the increase of the number of functions. On the contrary, for GCC, when the number of functions is small, its proportions is sufficiently low. It illustrates that the time consumed is longer for \emph{overheadTime} in GCC. When more functions participate into parallel workflows, \emph{funTime} of GCC gradually increases. The possible reason is that there are many parallel functions and the execution scheduling between them is heavy. 

\begin{figure}[htbp!]
  \centering
  \includegraphics[width=\linewidth]{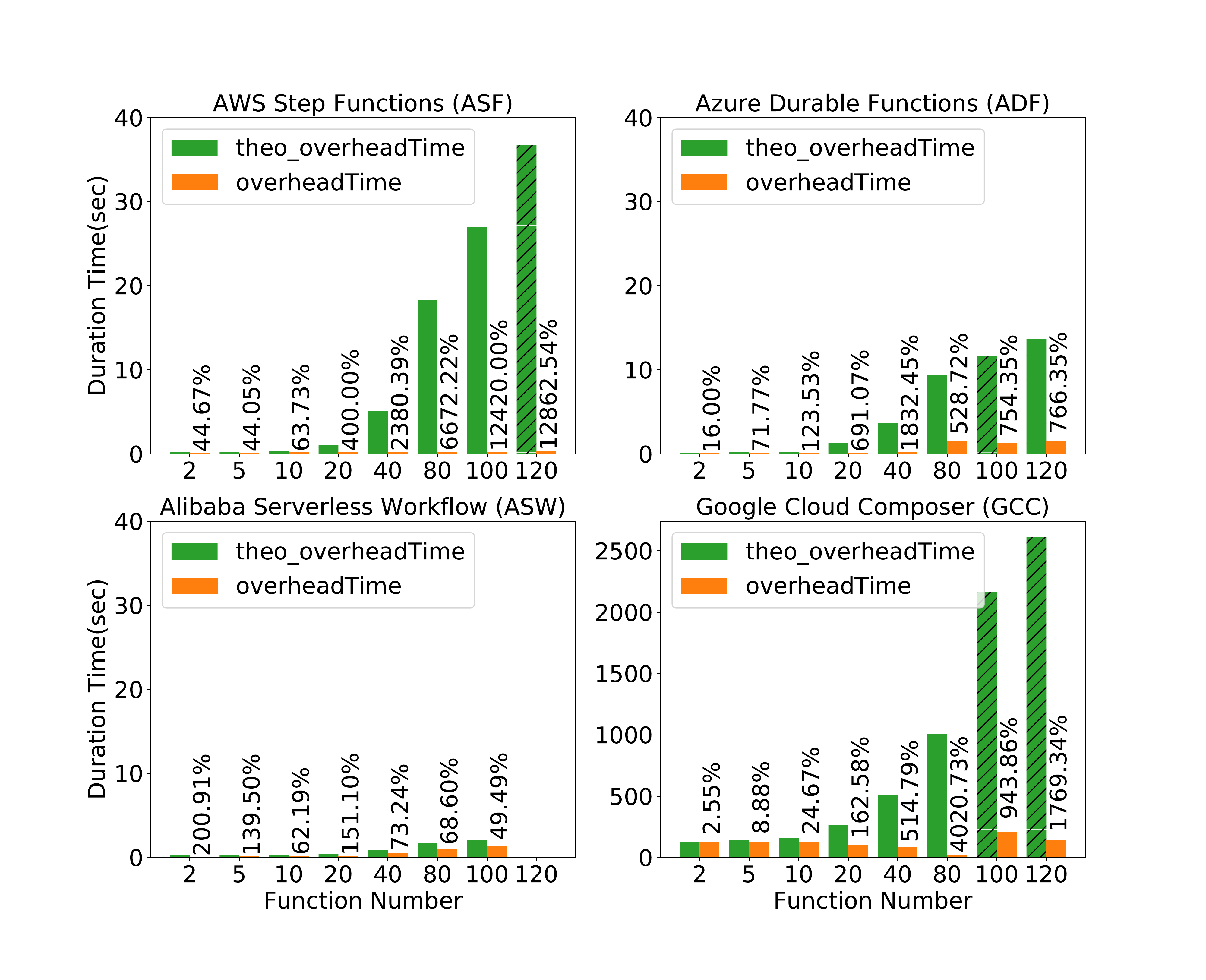}
  \caption{The comparison of \emph{overheadTime} and \emph{theo$\_$overheadTime} under various levels of activity complexity in parallel workflow.}
  \label{fig:FunctionNumPar-overheadTime+theoverheadTime}
\end{figure}

In parallel workflows, theoretically, all serverless functions with the same task start and complete at the same time. Thus, excluding the execution time of a single function from \emph{totalTime} is the theoretical orchestration overhead (\emph{theo$\_$overheadTime}) of this workflow. Figure~\ref{fig:FunctionNumPar-overheadTime+theoverheadTime} represents the comparison of \emph{overheadTime} and \emph{theo$\_$overheadTime} under various numbers of functions contained in parallel workflows for ASF, ADF, ASW, and GCC. It shows that \emph{theo$\_$overheadTime} increases as more functions are added into parallel workflows. We find that \emph{theo$\_$overheadTime} is much larger than \emph{overheadTime}. The value next to the bar indicates the percentage of \emph{theo$\_$overheadTime} higher than \emph{overheadTime}. Specifically, for ASF, its value can arrive as high as 12862.54$\%$, ADF is as high as 1832.45$\%$, ASW is as high as 200.91$\%$, and GCC is as high as 4020.73$\%$. In addition, as the number of parallel functions increases, \emph{overheadTime} becomes longer for ASF, ADF, and ASW. It takes a certain amount of time to process the branch and merge in parallel workflows. However, there is certain fluctuation in \emph{overheadTime} of GCC, and fluctuation may be caused by its environment.

\begin{figure}[ht]
  \centering
  \includegraphics[width=\linewidth]{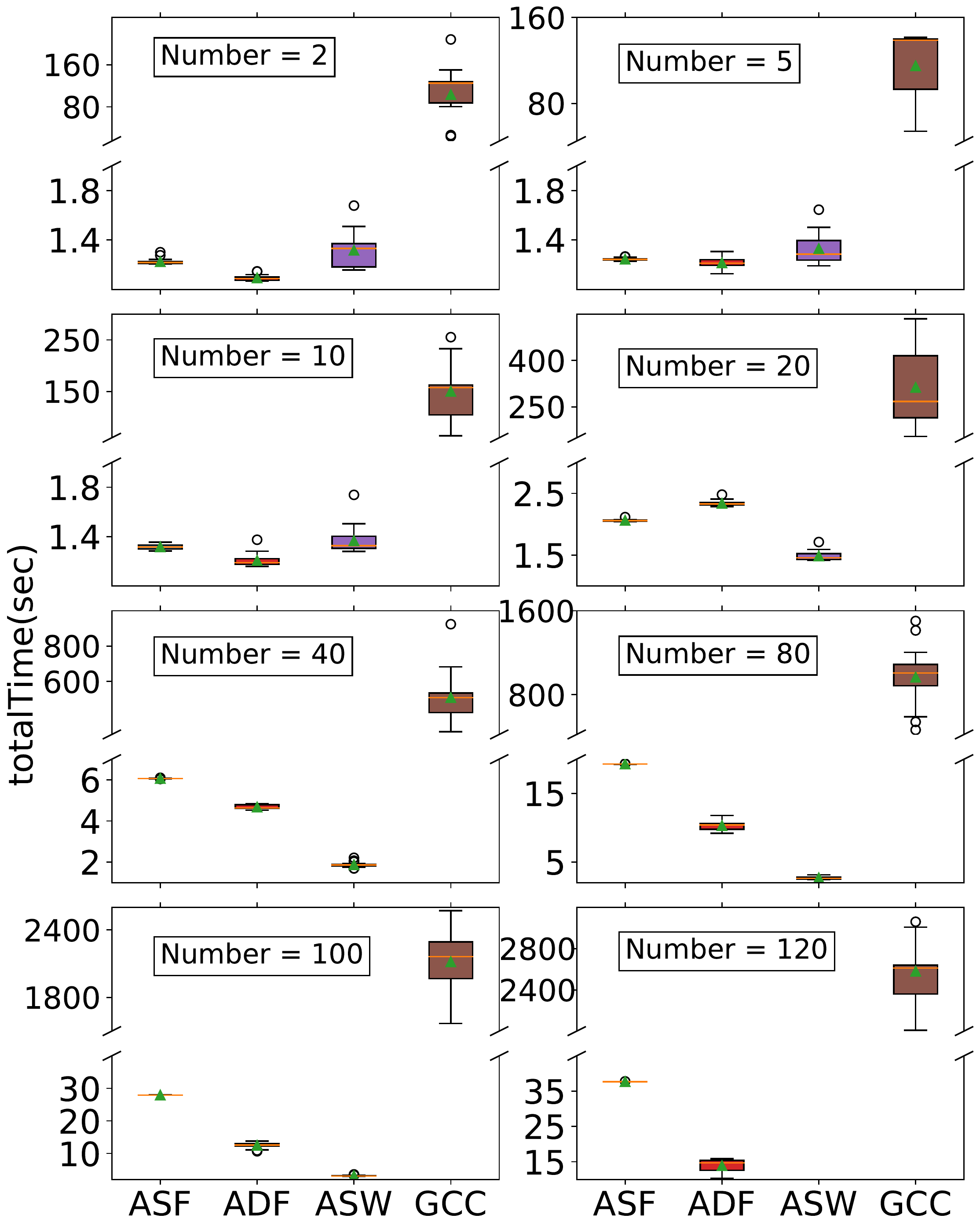}
  \caption{The comparison of \emph{totalTime} under various levels of activity complexity in parallel workflow.}
  \label{fig:FunctionNumPar-compare-totalTime}
\end{figure}

\begin{figure}[ht]
  \centering
  \includegraphics[width=\linewidth]{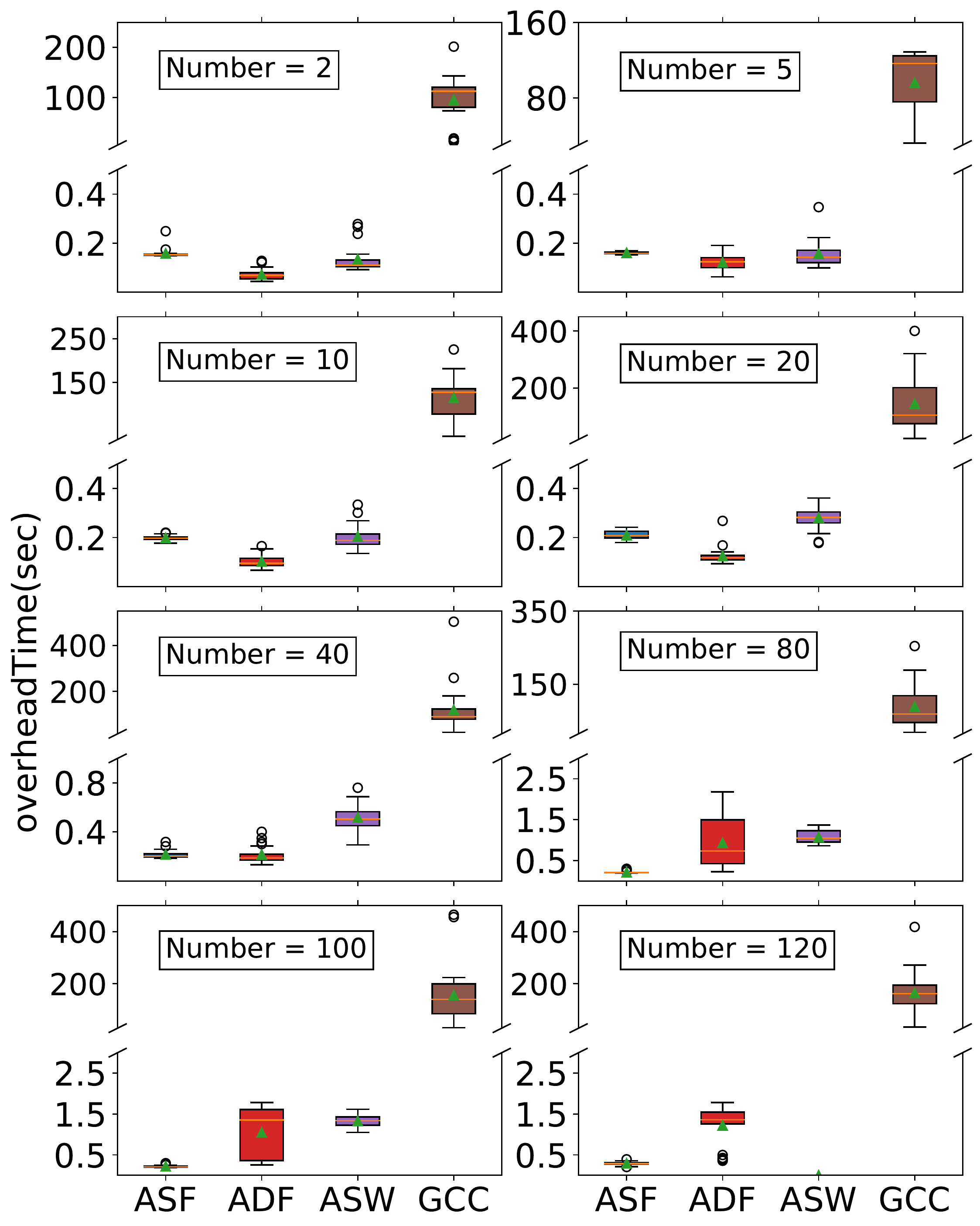}
  \caption{The comparison of \emph{overheadTime} under various levels of activity complexity in parallel workflow.}
  \label{fig:FunctionNumPar-compare-overheadTime}
\end{figure}

Figure~\ref{fig:FunctionNumPar-compare-totalTime} and Figure~\ref{fig:FunctionNumPar-compare-overheadTime} are comparisons of \emph{totalTime} and \emph{overheadTime} under various numbers of functions contained in parallel workflows. Figure~\ref{fig:FunctionNumPar-compare-totalTime} shows that GCC has the longest \emph{totalTime} in parallel experiments. When parallel functions with small-scale (less than or equal to 10), \emph{totalTime} of ASF, ADF, and ASW is not much different, but the result of ADF is the lowest. When more functions (between 10 and 100) are paralleled into the workflow, ASW begins to show its advantages that have the \emph{totalTime} result with lower and more stable compared to ASF and ADF. Since ASW has a limit of 100 for parallel tasks, no more parallel functions can be executed. In the case of parallel functions with greater than 100, ADF can complete workflows in a shorter\emph{totalTime}. Through observing the distribution of \emph{funTime}, we find that the changing characteristics of \emph{funTime} is the same as \emph{totalTime} of Figure~\ref{fig:FunctionNumPar-compare-totalTime} for ASF, ADF, and ASW. Due to space reasons, the distribution figure is not displayed. It illustrates that the changing trend of \emph{totalTime} depends on \emph{funTime}. Figure~\ref{fig:FunctionNumPar-overheadTime+theoverheadTime} shows that the number of functions affects \emph{overheadTime} of parallel workflows. Similarly, Figure~\ref{fig:FunctionNumPar-compare-overheadTime} also shows such characteristics. Moreover, when the number of functions does not exceed 40, ADF has the lowest \emph{overheadTime}. As the number of functions increases from 40 to 120, ASF exhibits lower \emph{overheadTime} than ADF, ASW, and GCC. However, \emph{overheadTime} values are relatively small for ASF, ADF, and ASW, and have little effect on their \emph{totalTime}. Thus, in actual scenarios, the effect of \emph{totalTime} in parallel workflows is usually considered. 

\textbf{For ES1, see Findings F.1, F.2, F.3, F.4 and Implications I.1, I.2, I.3, I.4 in Table~\ref{tab:findingandimplication}.}

\subsection{Data-flow Complexity (ES2)}

Data complexity reflects on the sizes of data payloads passed among serverless functions in a workflow.

\subsubsection{Sequence application scenario}

\begin{figure}[htbp!]
  \centering
  \includegraphics[width=\linewidth]{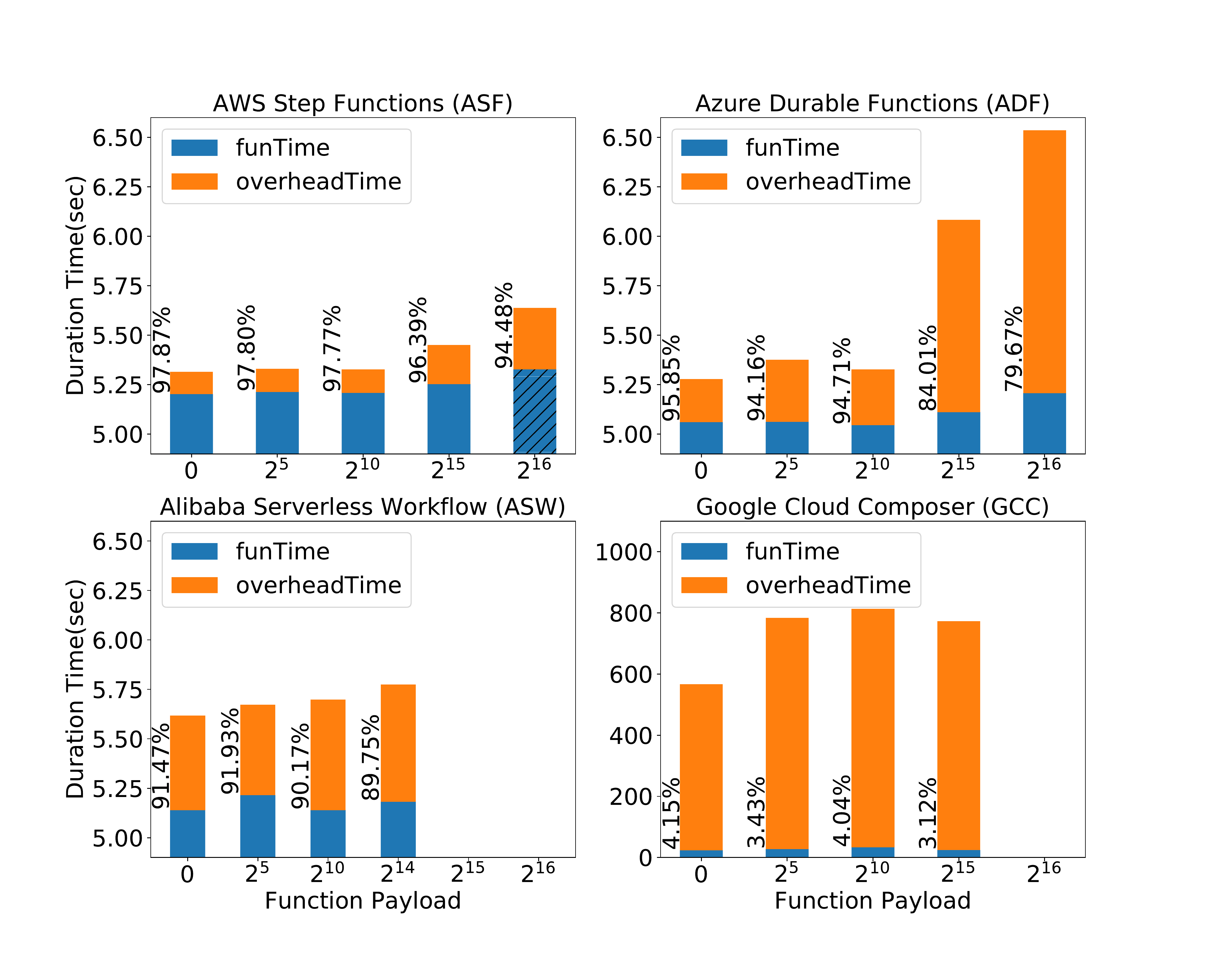}
  \caption{The performance of various levels of data-flow complexity in sequence workflow.}
  \label{fig:FunctionPayloadSeq-totalTime+funTime+overheadTime}
\end{figure}

Figure~\ref{fig:FunctionPayloadSeq-totalTime+funTime+overheadTime} shows the performance of data payloads between 0B to $2^{16}$B for ASF, ADF, ASW, and GCC. We add additional measurements for each serverless workflow service and find some problems. (i) For ASF, the size limit of the data payload in a workflow is 256KB ($2^{18}$B). We conduct measurements about the data payload with $2^{18}$B, and its \emph{totalTime}, \emph{funTime}, and \emph{overheadTime} are respectively 6,599s, 5,683, and 0.916s. In addition, when we add additional measurements that the data payload is large than $2^{18}$B, a validation error is detected, and it prompts the value at ``input'' failed to satisfy the constraint and must have a length less than or equal to 262,144 (i.e., $2^{18}$B). This error illustrates the data payload restriction described in the ASF document is consistent with the actual usage. (ii) For ADF, its document does not mention its size limit about the data payload. In order to measure whether ADF supports a larger data payload, we conduct measurements of the data payload with $2^{20}$B, and its \emph{totalTime}, \emph{funTime}, and \emph{overheadTime} are 27.613s, 6.688s, and 20.925s, respectively. (iii) For ASW, it exists the concept of the local variable. When the payload is set as $2^{15}$B, we find that a failure occurred. It also verifies that the total size of the input, output, and local variables of the step in ASW cannot exceed 32KB. To observe the impact of the data payload, we add measurements of the data payload with $2^{14}$B to Figure~\ref{fig:FunctionPayloadSeq-totalTime+funTime+overheadTime}. (iv) For GCC, when the data payload is set as $2^{16}$B, a ``mysql'' error occurs that storing a message is bigger than 65,535 bytes. We check the environment resources of GCC and find that Cloud SQL is used to store Airflow metadata to minimize the possibility of data loss. Thus, experiments of the data payload with $2^{16}$B cannot be performed due to the storage limitation.

Figure~\ref{fig:FunctionPayloadSeq-totalTime+funTime+overheadTime} shows that when the data payload is less than or equal to $2^{10}$B, \emph{totalTime}, \emph{funTime}, and \emph{overheadTime} of ASF, ADF, and ASW have little effect. When the data payload is greater than $2^{10}$B, \emph{totalTime} of ASF and ASW increases slightly. However, considering the results of ADF in the data payload $2^{20}$B, we find that \emph{totalTime} of ADF increases significantly. Thus, \emph{we conclude that the ASF, ADF, and ASW have a little impact under low data payload conditions. Only under high data payload conditions will ASF, ADF, and ASW have a certain impact.} We also find that \emph{totalTime} about the data payload $2^{16}$B of ADF is not much different from \emph{totalTime} about the data payload $2^{18}$B of ASF. It illustrates that ASF is more suitable in high data payloads (between $2^{10}$B and $2^{18}$B). However, ADF can achieve larger data payloads (larger than $2^{18}$B) in sequence workflow than ASF, ASW, and GCC. For GCC in Figure~\ref{fig:FunctionPayloadSeq-totalTime+funTime+overheadTime}, we observe that the data payload has much effect on \emph{overheadTime} compared with  \emph{funTime}. Similarly, it shows a consistent conclusion with ``F.2'' of Table~\ref{tab:findingandimplication}.

\begin{figure}[ht]
  \centering
  \includegraphics[width=\linewidth]{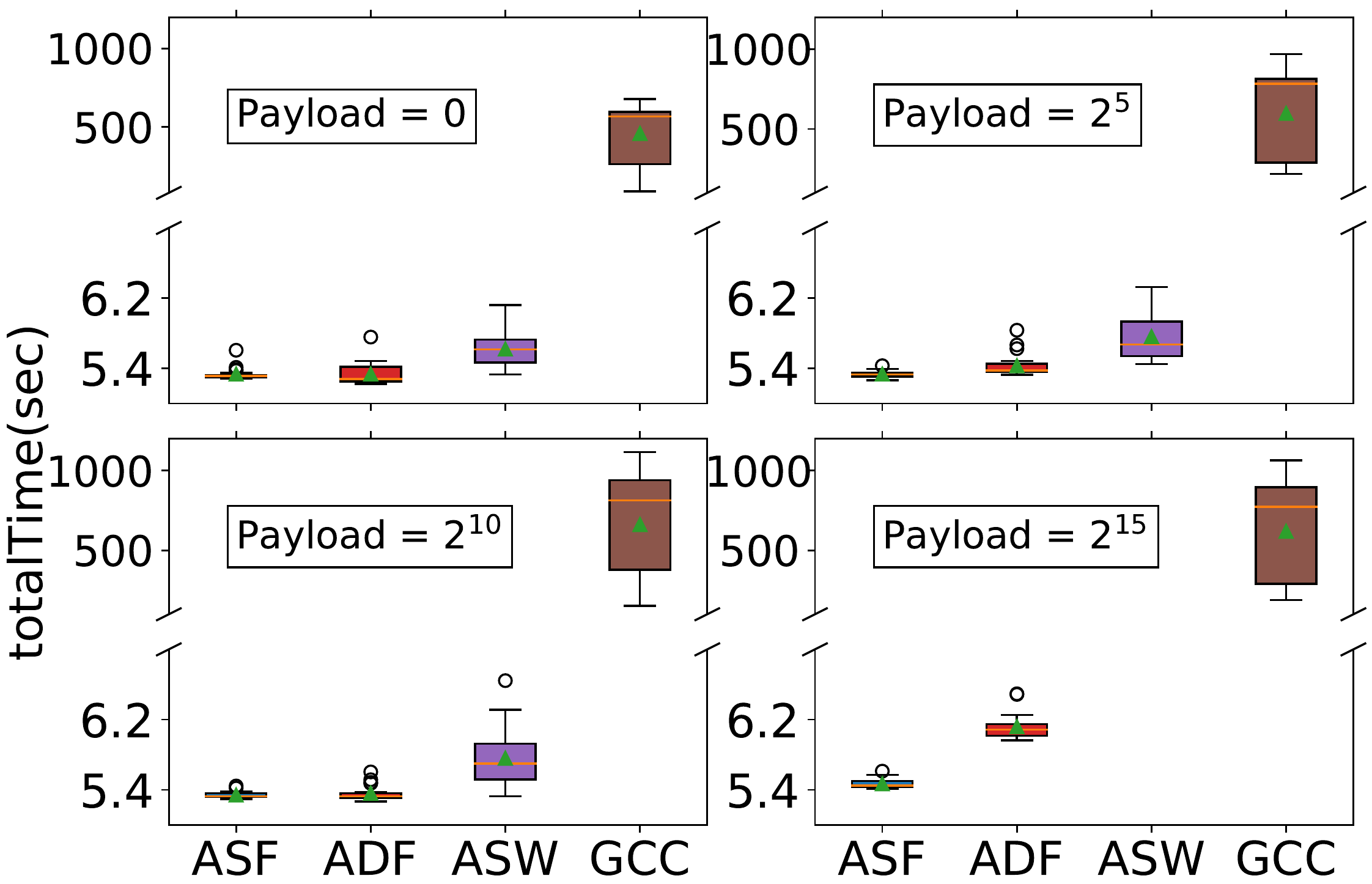}
  \caption{The comparison of \emph{totalTime} under various levels of data-flow complexity in sequence workflow.}
  \label{fig:FunctionPayloadSeq-compare-totalTime}
\end{figure}

\begin{figure}[ht]
  \centering
  \includegraphics[width=\linewidth]{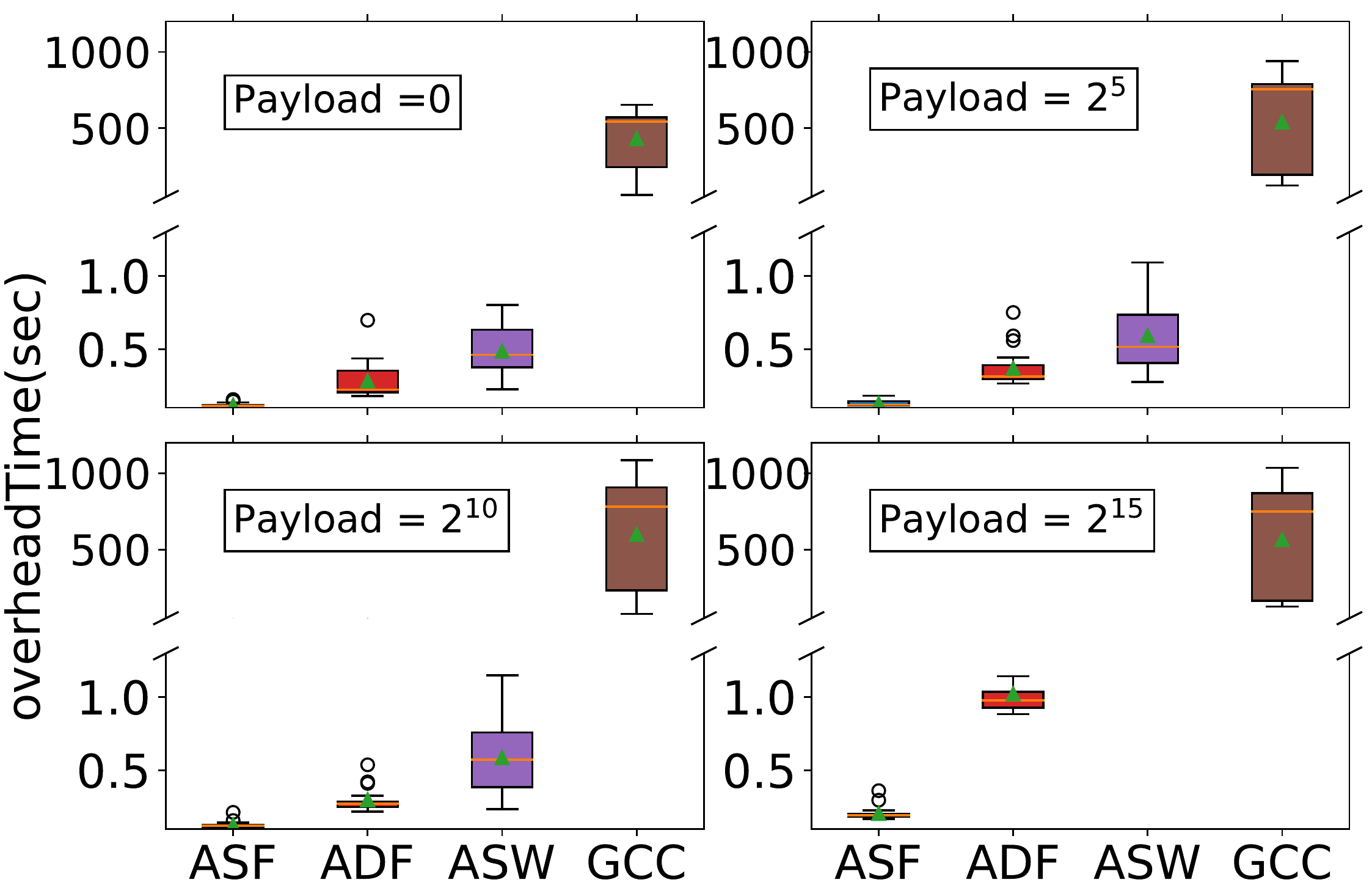}
  \caption{The comparison of \emph{overheadTime} under various levels of data-flow complexity in sequence workflow.}
  \label{fig:FunctionPayloadSeq-compare-overheadTime}
\end{figure}

To compare the result distribution of measurements, we draw their respective box plots. Figure~\ref{fig:FunctionPayloadSeq-compare-totalTime} shows that the comparison of \emph{totalTime} under various data payloads in sequence workflows. In the low data payload range ( $\leq$ $2^{10}$B), \emph{totalTime} of ASF, ADF, and ASW is not much different, and \emph{totalTime} of ASF and ADF is lower than ASW. In the high data payload range (between $2^{10}$B and $2^{15}$B), \emph{totalTime} of ASF is the lowest. Considering previous analysis about the data payload with between $2^{15}$B and $2^{18}$B in Figure~\ref{fig:FunctionPayloadSeq-totalTime+funTime+overheadTime}, similarly, ASF is lower than ADF with regard to \emph{totalTime}. However, whether in the low data payload or high data payload, \emph{totalTime} of GCC is the highest. For the distribution of \emph{funTime}, similar to Figure~\ref{fig:FunctionNumSeq-compare-funTime}, ADF has the lowest \emph{funTime}, followed by ASW, ASF, and finally GCC. Due to space reasons, the distribution figure is not displayed. Specifically, when payloads are within $2^{10}$B, values of ASF, ADF, and ASW maintain between 5s and 5.6s, while GCC is between 20s and 150s. Figure~\ref{fig:FunctionPayloadSeq-compare-overheadTime} is the comparison of \emph{overheadTime} under various data payloads in sequence workflows. We observe that \emph{overheadTime} of ASF is the lowest among them, and its result distribution is more compact and stable when the data payload is no greater than $2^{15}$B. Additionally, previous results of the data payload with between $2^{15}$B and $2^{16}$B in Figure~\ref{fig:FunctionPayloadSeq-totalTime+funTime+overheadTime} also shows that ASF is lower than ADF with regard to \emph{overheadTime}. In this situation, ASF is the most best choose. However, when the data payload passing among functions grows to over 256KB, if developers still want to use ASF, advice to use Amazon S3 to store the data, and pass the Amazon Resource Name (ARN) instead of raw data.

\subsubsection{Parallel application scenario}

\begin{figure}[htbp!]
  \centering
  \includegraphics[width=\linewidth]{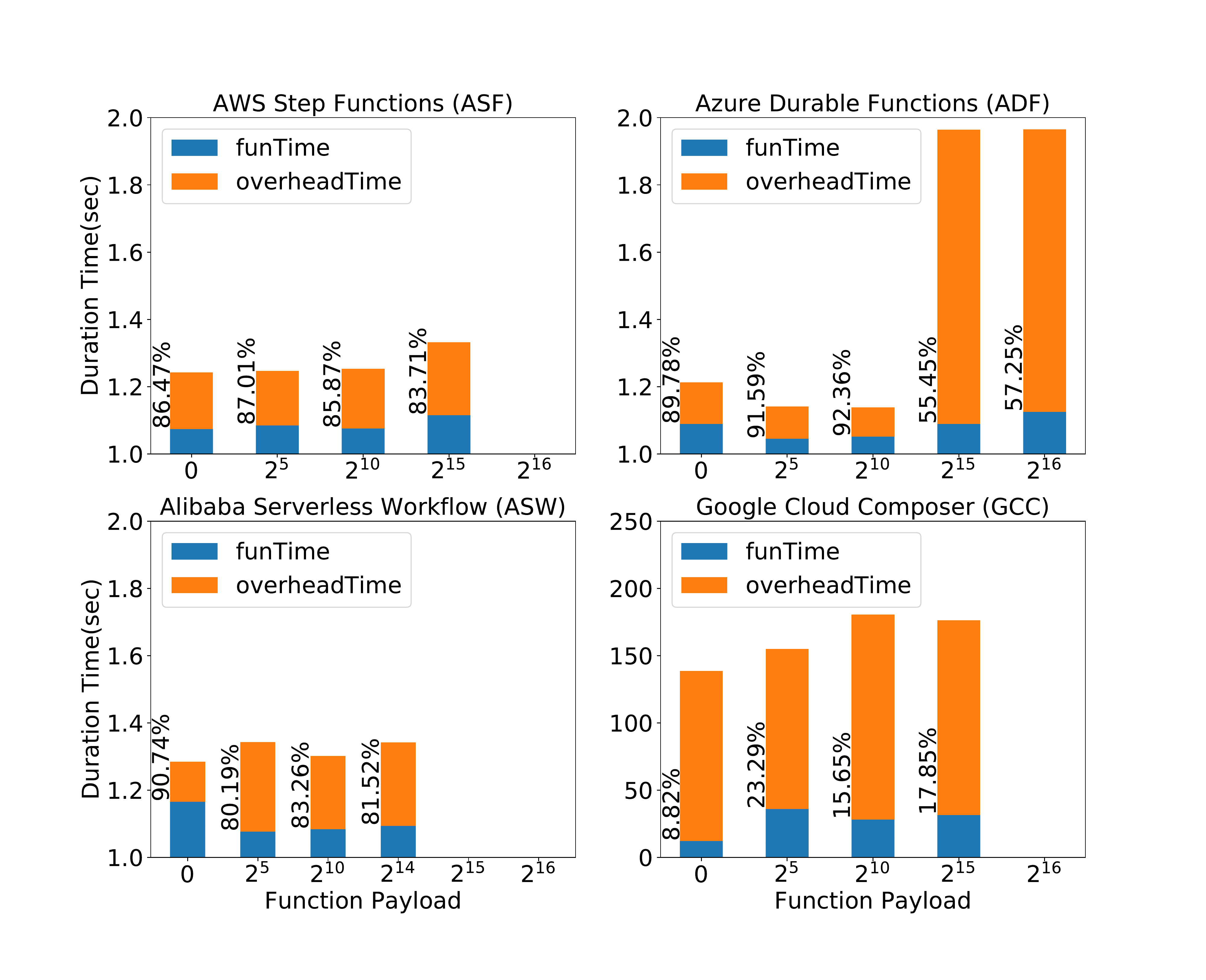}
  \caption{The performance of various levels of data-flow complexity in parallel workflow.}
  \label{fig:FunctionPayloadPar-totalTime+funTime+overheadTime}
\end{figure}

Figure~\ref{fig:FunctionPayloadPar-totalTime+funTime+overheadTime} represents the performance of various data payloads in parallel workflow for ASF, ADF, ASW, and GCC. First, we discuss the performance of \emph{totalTime}. \emph{totalTime} of ASF and ASW is basically not affected by low data payloads. However, transmit the data payload with $2^{16}$B into parallel workflows in ASF, and produce 5 times $2^{16}$B data size to the workflow output. The high data output (larger than $2^{18}$B) causes a failure of the workflow execution. For ASW, it has the data payload limit (32KB), and the merge of parallel functions also need to be considered. From Figure~\ref{fig:FunctionPayloadPar-totalTime+funTime+overheadTime}, we observe that when data payloads are within $2^{10}$B in ADF, its \emph{totalTime} keeps stable. In the high data payload (> $2^{10}$B), \emph{totalTime} of ADF increases. We also carry out the parallel experiments with a data payload of $2^{20}$B, where \emph{totalTime} is 2,130s that is larger than the data payload with $2^{16}$B. For GCC, \emph{totalTime} is affected by whether there is a payload or not. When there is a payload, \emph{totalTime} will increase, but as the payload increases, it does not show a regular trend. Then, we discuss the performance of \emph{funTime} and \emph{overheadTime}. \emph{funTime} and \emph{overheadTime} of ASF and ASW do not change much and are basically stable (maintain acceptable fluctuations, e.g., 100ms). For ADF, under high data payloads, \emph{overheadTime} increases greatly, and \emph{funTime} does not change much. Thus, \emph{overheadTime} of ADF under high data payloads is the main reason affecting \emph{totalTime} change in parallel workflows. For GCC, \emph{funTime} and \emph{overheadTime} both increase in parallel workflow with the payload transmission. Thus, \emph{only under high payload conditions will ASF, ADF, SW have a certain impact, while GCC is affected by whether there is a payload or not.} 

\begin{figure}[ht]
  \centering
  \includegraphics[width=\linewidth]{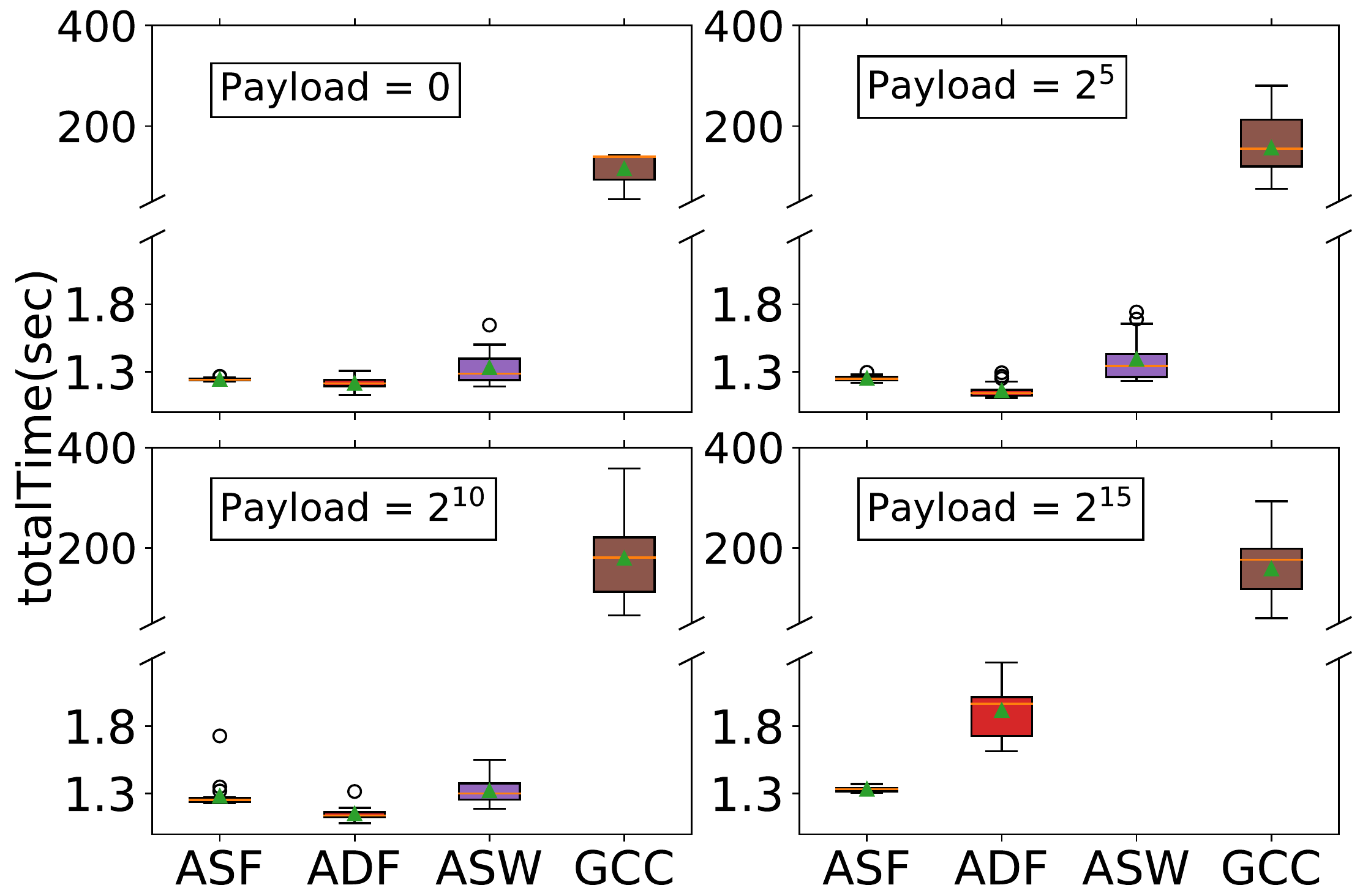}
  \caption{The comparison of \emph{totalTime} under various levels of data-flow complexity in parallel workflow.}
  \label{fig:FunctionPayloadPar-compare-totalTime}
\end{figure}

The result distribution of \emph{totalTime} is shown in Figure~\ref{fig:FunctionPayloadPar-compare-totalTime}. Result distributions of \emph{funTime} and \emph{overheadTime} are similar to Figure~\ref{fig:FunctionPayloadPar-compare-totalTime}. Due to space reasons, figures are not displayed. When the data payload is set to be small ($\leq$ $2^{10}$B), \emph{totalTime}, \emph{funTime}, and \emph{overheadTime} of ASF and ADF are low and relatively stable, whereas GCC is discrete and volatile. When data payloads are between $2^{10}$B and $2^{15}$B, \emph{totalTime}, \emph{funTime}, and \emph{overheadTime} of ASF are lowest. However, if developers want to pass into a large payload, only ADF or ASF with the external storage can execute in parallel workflows. 

\textbf{For ES2, see Findings I.5, I.6, I.7 and Implications I.5, I.6, I.7 in Table~\ref{tab:findingandimplication}.}

\subsection{Function Complexity (ES3)}

Function complexity reflects on the specified duration time of serverless functions contained in a workflow. 

\subsubsection{Sequence application scenario}

\begin{figure}[htbp!]
  \centering
  \includegraphics[width=\linewidth]{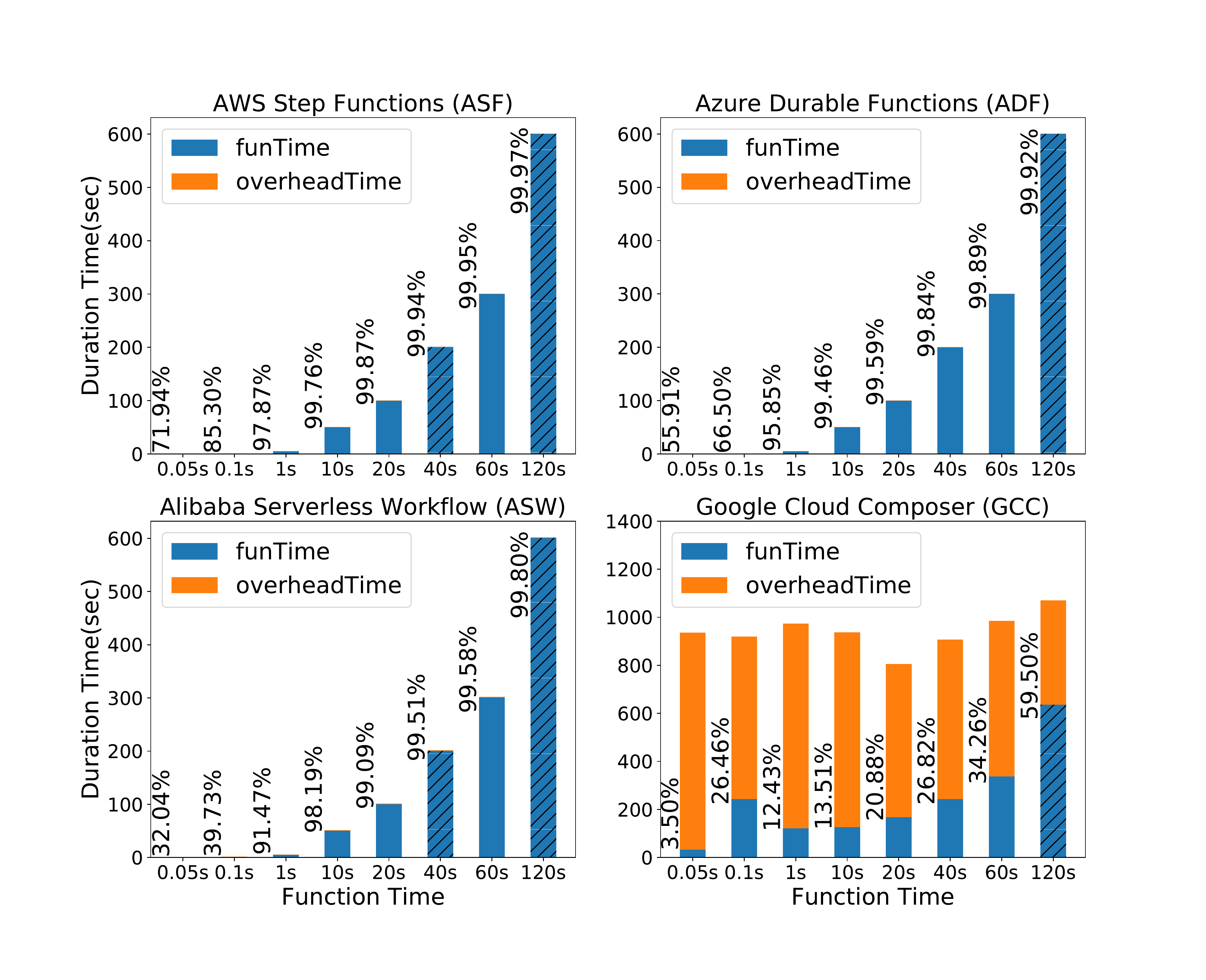}
  \caption{The performance of various levels of function complexity in sequence workflow.}
  \label{fig:FunctionTimeSeq-totalTime+funTime+overheadTime}
\end{figure}

Figure~\ref{fig:FunctionTimeSeq-totalTime+funTime+overheadTime} represents the performance of various specified duration times of functions in sequence workflows for ASF, ADF, ASW, and GCC. \emph{totalTime} of ASF, ADF, and ASW increases as the specified duration time of functions gradually grows, whereas \emph{totalTime} of GCC is not affected. Besides, there is no obvious trend in GCC for \emph{funTime} and \emph{overheadTime} under various specified duration times of functions. This situation in Figure~\ref{fig:FunctionTimeSeq-totalTime+funTime+overheadTime} is as described in ``F.2'' of Table~\ref{tab:findingandimplication}.

The specific changes about \emph{overheadTime} are shown in Table~\ref{tab:overheadtimesequence}. When the number of functions contained in sequence workflows is fixed, \emph{overheadTime} generally does not increase significantly for ASF, ADF, ASW, and they are roughly maintained within a certain range. 

For ASF, the range is 0.1s to 0.2s, ADF is 0.2s to 0.5s, and ASW is 0.5s to 1.2s. Thus, we conclude that \emph{changes within the function may not affect \emph{overheadTime}, whereas changes between the workflow structure and data payload may have a certain impact on it.} However, the fluctuation of \emph{overheadTime} of GCC is relatively large, ranging from 430s to 900s.

\begin{table}[htb]
 \centering
 \small
 \caption{Median of \emph{overheadTime} (seconds) about various specified duration times of functions in sequence workflow.}
 \begin{adjustbox}{width=1\columnwidth}
    \begin{tabular}{|p{0.1\columnwidth}|p{0.1\columnwidth}|p{0.1\columnwidth}|p{0.1\columnwidth}|p{0.1\columnwidth}|p{0.1\columnwidth}|p{0.1\columnwidth}|p{0.1\columnwidth}|p{0.1\columnwidth}|}
\hline

\rowcolor{gray!50}
	& \textbf{0.05s}  & \textbf{0.1s}  & \textbf{1s}  & \textbf{10s}  & \textbf{20s}  & \textbf{40s}  & \textbf{60s}  & \textbf{120s}  \\
\hline

\cellcolor{gray!15}\textbf{ASF}	& 0.188 & 0.121 & 0.113 & 0.119 & 0.127 & 0.120 & 0.150 & 0.157 \\
\hline

\cellcolor{gray!15}\textbf{ADF}	& 0.246 & 0.272 & 0.219 & 0.271 & 0.408 & 0.320 & 0.339 & 0.507 \\
\hline

\cellcolor{gray!15}\textbf{ASW}	    & 0.738 & 1.036 & 0.479 & 0.926 & 0.922 & 0.978 & 1.264 & 1.207 \\
\hline

\cellcolor{gray!15}\textbf{GCC}	& 903.720 & 676.320 & 852.360 & 810.600 & 637.020 & 663.540 & 647.280 & 433.320 \\
\hline
\end{tabular}
\end{adjustbox}
\label{tab:overheadtimesequence}
\end{table}

Due to space reasons, distribution figures about \emph{totalTime}, \emph{funTime}, and \emph{overheadTime} are not displayed. We find that both ASF and ADF have lower \emph{totalTime} than ASW and GCC. Furthermore, the measurement results of ASF are more stable than ADF. For the \emph{funTime} distribution, similar to Figure~\ref{fig:FunctionNumSeq-compare-funTime}, ADF has the lowest \emph{funTime}, followed by ASW, ASF, and finally GCC. Regarding the distribution of \emph{overheadTime}, we find that ASF has the lowest result among all serverless workflow services overall.

\subsubsection{Parallel application scenario}

\begin{figure}[htbp!]
  \centering
  \includegraphics[width=\linewidth]{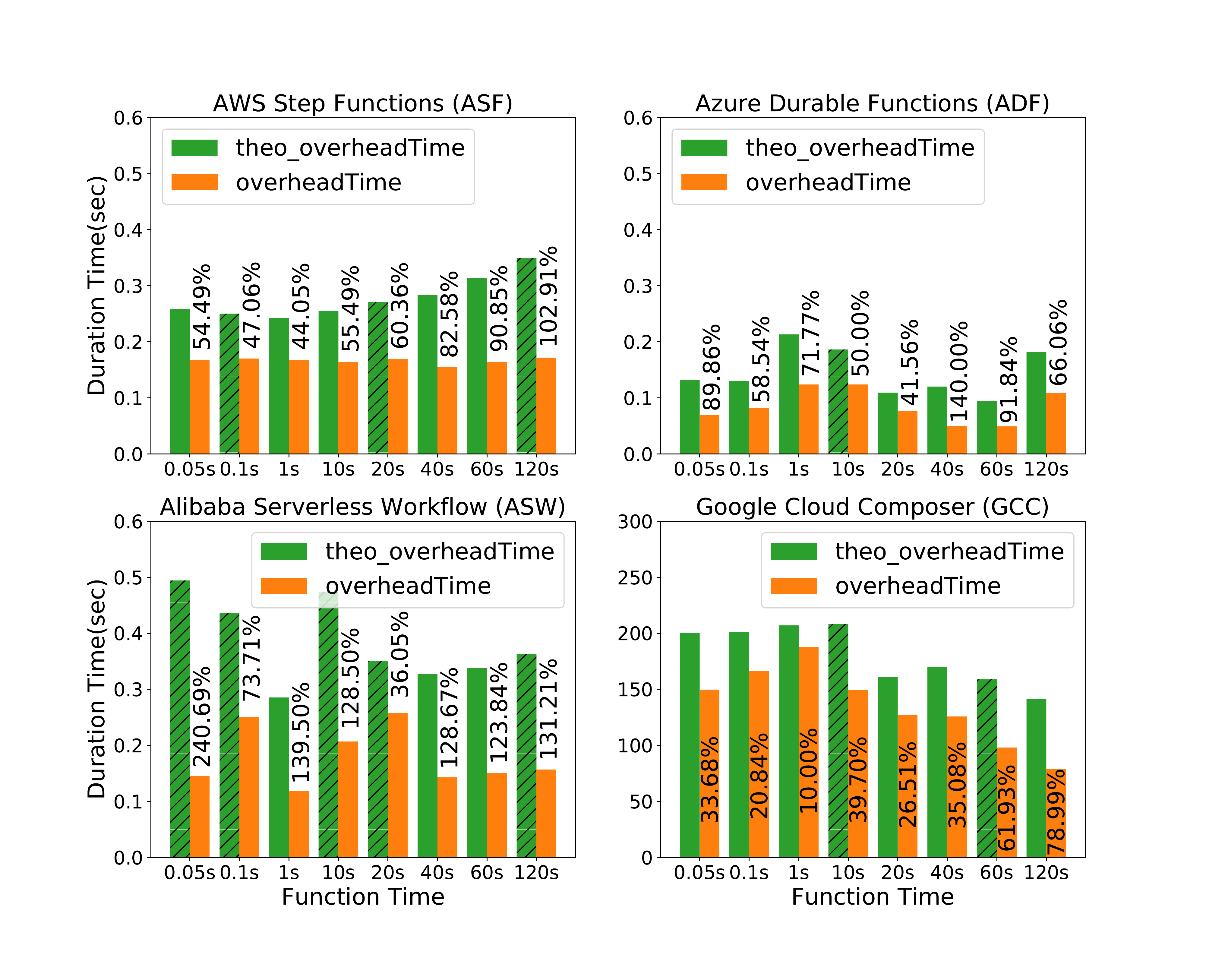}
  \caption{The comparison  of \emph{overheadTime} and \emph{theo$\_$overheadTime} under various levels of function complexity in parallel workflow.}
  \label{fig:FunctionTimePar-overheadTime+theoverheadTime}
\end{figure}

When the number of functions contained in parallel workflows is deterministic and the specified duration time of functions increases, \emph{totalTime} and \emph{funTime} must increase. However, \emph{totalTime} and \emph{funTime} of GCC does not increase with the increase of the specified duration time of functions in parallel workflows. Due to space reasons, this figure is not displayed. To observe the changes in \emph{overheadTime} more intuitively, Figure~\ref{fig:FunctionTimePar-overheadTime+theoverheadTime} shows the comparison for \emph{overheadTime} and \emph{theo$\_$overheadTime}. We find that \emph{theo$\_$overheadTime} is larger than \emph{overheadTime}. We also find that \emph{overheadTime} and \emph{theo$\_$overheadTime} for ASF, ADF, ASW, and GCC do not change significantly with the growth of the specified duration time of functions, and they basically fluctuate within a certain range. ASF, ADF, and SW are below 0.5s, while GCC is below 200s. At the same time, Figure~\ref{fig:FunctionTimePar-overheadTime+theoverheadTime} shows that \emph{overheadTime} of ASF is relatively stable, whereas ADF, ASW, and GCC have certain fluctuations.

The distributions of \emph{totalTime}, \emph{funTime}, and \emph{overheadTime} are similar. Due to space reasons, figures are not displayed. We find that the results of ADF are the lowest in terms of \emph{totalTime}, \emph{funTime}, and \emph{overheadTime}, whereas GCC is the highest. However, as far as the stability of the result is concerned, ASF is the best. For ASW, its results are relatively unstable compared with ADF, and there are more outliers.

\textbf{For ES3, see Findings F.8, F.9 and Implications I.8, I.9 in Table~\ref{tab:findingandimplication}.}

\section{Discussion}\label{discussion}

For verifying our findings, we conduct experiments of two serverless application workloads, i.e., KMeans and MapReduce. Then, we discuss some limitations of our study.

\noindent\textbf{KMeans application} is implemented in a sequence workflow, and accomplishes the clustering functionality for point sets with three-dimensional space. First, use a serverless function to generate 1500 points, because the data payload limit of ASW cannot generate the data of 2000 points. Second, initialize the centroid points randomly. For the \emph{KMeans} algorithm, the K-value of clustering needs to be given in advance. We adopt \emph{Elbow Method} presented by Yuan \textit{et al.}~\cite{yuankmeansresearch2019} to determine K as 8. Next, based on the point set and centroid points, perform the clustering functionality of \emph{KMeans}. Finally, output and show the clustering result.

\begin{figure}[ht]
  \centering
  \includegraphics[width=\linewidth]{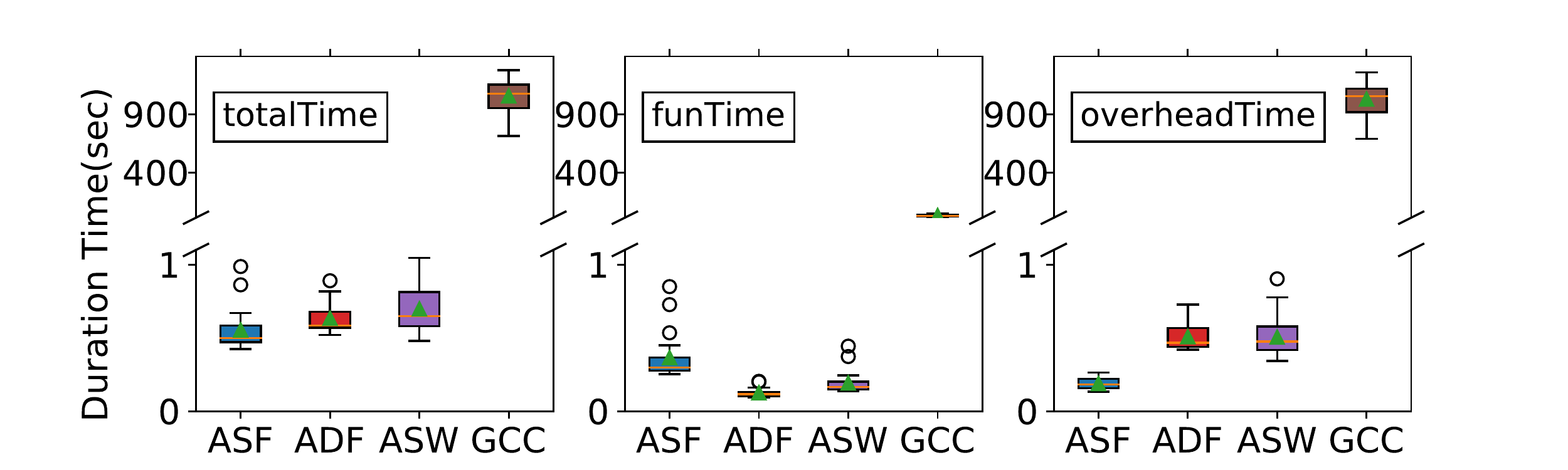}
  \caption{The performance of the \emph{KMeans} application.}
  \label{fig:KMeans-compare-totalTime+funTime+overheadTime}
\end{figure}

\begin{figure}[ht]
  \centering
  \includegraphics[width=\linewidth]{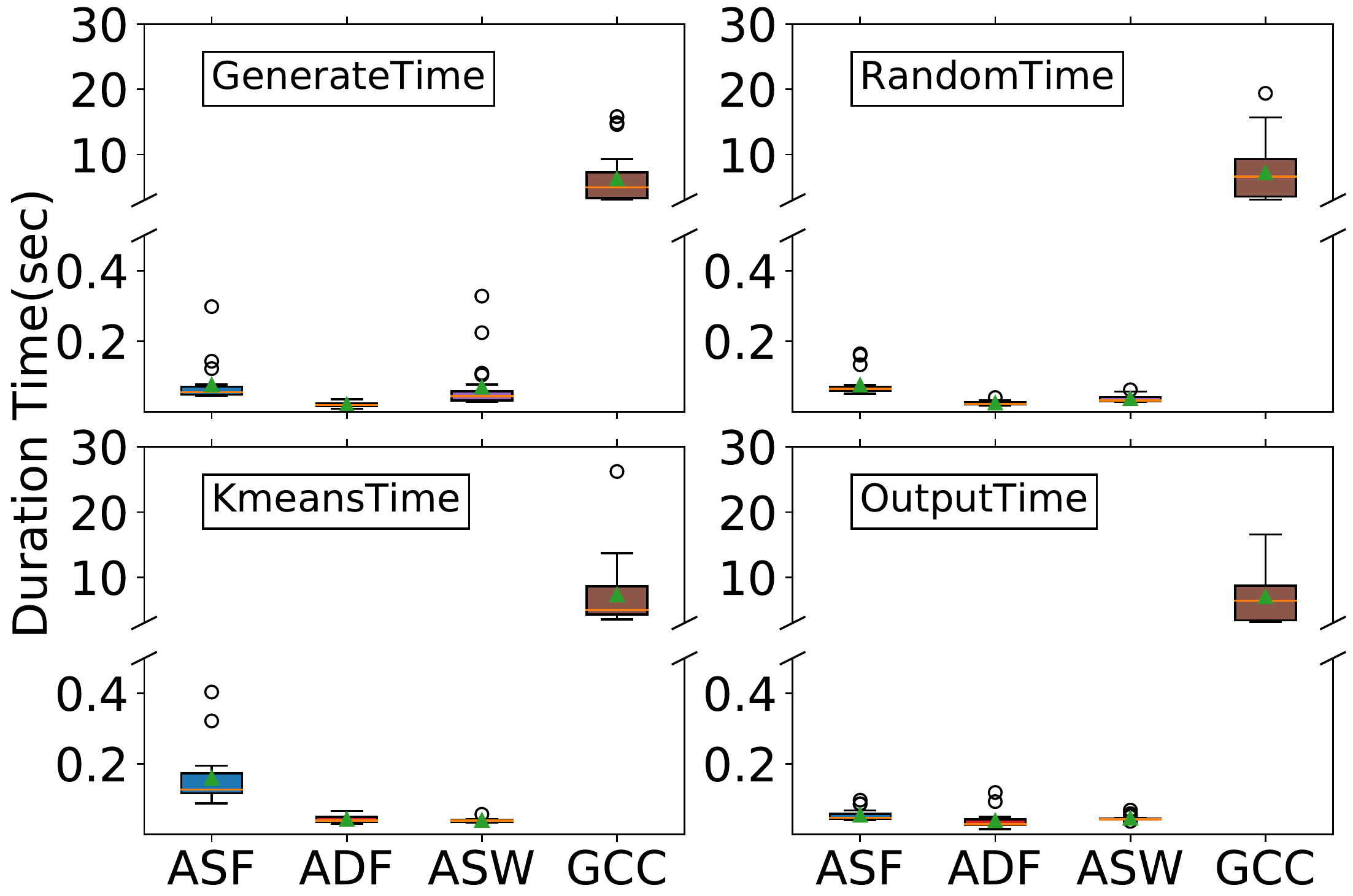}
  \caption{The comparison about execution times of respective functions about the \emph{KMeans} application.}
  \label{fig:KMeans-compare-function}
  \vspace{-0.1cm}
\end{figure}

Figure~\ref{fig:KMeans-compare-totalTime+funTime+overheadTime} represents the comparison of \emph{totalTime}, \emph{funTime}, and \emph{overheadTime} of the \emph{KMeans} application for ASF, ADF, ASW, and GCC. ASF shows the shortest \emph{totalTime} and \emph{overheadTime}, and ADF has the shortest \emph{funTime} (\textbf{F.8 in Table~\ref{tab:findingandimplication}}). This is consistent with the implication \textbf{I.3 in Table~\ref{tab:findingandimplication}}. We also find the same inferred that the changing trend of \emph{totalTime} in sequence workflow is mainly affected by \emph{overheadTime} (\textbf{F.3 in Table~\ref{tab:findingandimplication}}) because \emph{funTime} cost the relatively low and stable time in this \emph{KMeans} application. In terms of data-flow complexity about the data payloads, the previous conclusion (\textbf{I.6 in Table~\ref{tab:findingandimplication}}) is that when the data payload is less than $2^{18}$, ASF is advised to use. In the \emph{KMeans} application, the data payload size is within $2^{18}$, and the performance of ASF is best considering \emph{totalTime}, \emph{overheadTime}. Figure~\ref{fig:KMeans-compare-function} shows execution times of respective functions. Compared with GCC, the execution time of each function of ASF, ADF, and ASW is lower and more stable. It can still be concluded that the performance ADF is the best on the function execution (\textbf{F.8 in Table~\ref{tab:findingandimplication}}).

\noindent\textbf{MapReduce application} is implemented in a parallel workflow and is accomplished by the workflow solution example~\cite{MapReduceExample}. The application goal is to generate a batch of data to be processed, the value of data is $data\_1$ or $date\_2$. Count the number of occurrences of various data leveraging \emph{MapReduce} processing frame mode.

\begin{figure}[ht]
  \centering
  \includegraphics[width=\linewidth]{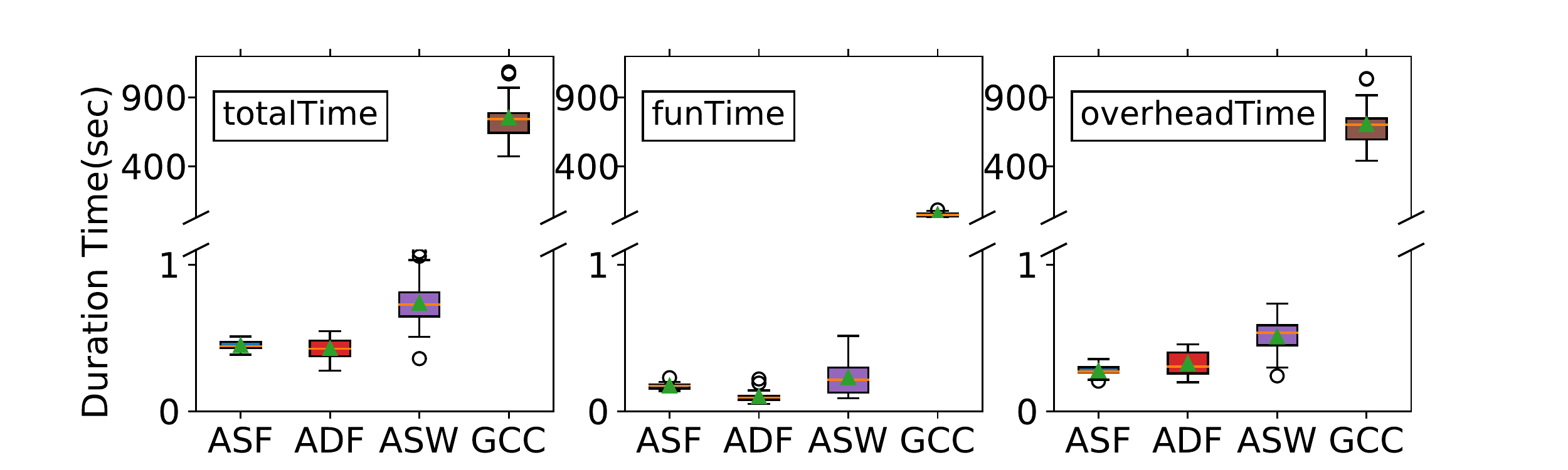}
  \caption{The performance of the \emph{MapReduce} application..}
  \label{fig:MapReduce-compare-totalTime+funTime+overheadTime}
\end{figure}

\textbf{I.1 in Table~\ref{tab:findingandimplication}} presents that ADF is used in small-scale activity-intensive parallel workflows. In Figure~\ref{fig:MapReduce-compare-totalTime+funTime+overheadTime}, it also shows ADF has the relatively short \emph{totalTime}. However, results from \emph{totalTime} and \emph{overheadTime} of ASF are more stable than ADF. In the \emph{MapReduce} application, there are certain data payload to be transmitted. In the presence of the data payload, the previous conclusion is the ASF is more suitable when data payloads are less than $2^{15}$B in parallel workflow (\textbf{I.6 in Table~\ref{tab:findingandimplication}}). Thus, results of ASF show a relatively satisfactory \emph{totalTime} and \emph{overheadTime}. For respective function execution times, we also find that the performance of ADF is best and the same as our previous conclusions (\textbf{F.8 in Table~\ref{tab:findingandimplication}}). Due to space reasons, the distribution figure about respective execution times of functions is not displayed.

\noindent\textbf{Limitations.} We discuss the limitations of our study. (i) \textbf{Selection of application scenarios.} Our study is based on sequence and parallel workflows. We may ignore other complex structures, e.g., choice, missing valuable insights with regard to the structure complexity of workflows. In future work, we plan to extend our study to diversify workflow structure to further obtain interesting findings. (ii) \textbf{Experiments of GCC.} In our study, the results of GCC fluctuate greatly, and we suppose it may be related to its environmental setting. To minimize this impact, we repeat several measurements. From the perspective of serverless computing, we suppose that functions performed in DAGs of GCC are not serverless, i.e., GCC is not designed for orchestrating serverless
functions. Until October 2020, we verify our assumption. Beta launch of \emph{Workflows}~\cite{GoogleWorkflow} service is released in the category of serverless computing of Google Cloud. However, its functionality has been immature yet, and we look forward to furthering research in our future work.

\section{Related Work}\label{related}

In this section, we summarize the related work serverless computing and serverless workflow. 

Serverless computing is a new paradigm of cloud computing. In general, computation offloading for various applications can be accomplished on cloud~\cite{add3,add14,add7}, mobile client~\cite{add12,add11}) using different hardware (such as CPU, GPU). Particularly, edge computing is an emerging and promising technology dedicated for improving use experience of today's interactive applications~\cite{add1}. This also promote the development of service usage. In this situation, service-oriented situational applications have shown great potential in solving immediate and quick roll-out problems~\cite{add4}. The concept and technique~\cite{add6,add5} of service computing is becoming more and more mature. Thus, Function-as-a-Service becomes a popular trend. Major cloud vendors present the correspond strategies like resource management rethink~\cite{add2,add9,add10} and performance~\cite{add13,add8} of previous mobile web work. In serverless computing, cloud paltforms uniformly manage resources to ensure scalability and load balancing.

Nowadays, serverless computing has already been used in various scenarios including Internet of Things and edge computing~\cite{de2016hierarchical}, data processing~\cite{jonas2017occupy, chard2017ripple, fouladi2017encoding}, scientific workflow~\cite{Malawski2020}, system security~\cite{bila2017leveraging}, etc. Authors generally believe running applications in a serverless architecture is more cost-efficient than microservices or monoliths. Wang \textit{et al.}~\cite{wang2018peeking} conducted the largest measurement study for AWS Lambda, Azure Functions, and Google Cloud Functions, and they used more than 50,000 function instances to characterize architectures, performance, and resource management. Preliminary measurements on AWS Lambda, Azure Functions, Google Cloud Functions, and IBM OpenWhisk were accomplished by McGrath \textit{et al.}~\cite{mcgrath2017serverless}, and they found AWS can achieve better scalability, cold-start latency, and throughput than other platforms.

To facilitate the coordination of serverless functions of applications, major cloud providers have proposed serverless workflow services. Authors~\cite{lopez2018comparison} compared and analyzed the serverless workflow services from their architectures, and programming in 2018. However, new information has been updated on the official website. For example, Azure Durable Functions currently has supported \emph{JavaScript} and \emph{Python} languages. Akkus \textit{et al.}~\cite{akkus2018sand} ran an image processing pipeline using serverless workflow services. They found that the total execution time of workflows is significantly more than the actual time required for function execution. As confirmed by our study, a part of the time overhead will be used for orchestration scheduling between serverless functions. Recently, Authors~\cite{lopez2020triggerflow} presented \emph{Triggerflow}, which is an extensible trigger-based orchestration architecture integrating various workflow models, e.g., \emph{State Machines}, \emph{Directed Acyclic Graphs}, and \emph{Workflow as code}. It can be seen that there is a growing interest in serverless workflow. Furthermore, we notice that serverless workflow services are evolving quickly. Nevertheless, our findings and implications serve as a snapshot of orchestration mechanisms, provide performance baselines and suggestions for developers to accomplish reliable and satisfying applications, and help cloud providers improve service efficiency.

\section{Conclusion}\label{conclusion}

In this paper, we present the first empirical study on characterizing and comparing existing serverless workflow services, i.e., AWS Step Functions, Azure Durable Functions, Alibaba Serverless Workflow, and Google Cloud Composer. We first compare their characteristics from six dimensions, e.g., orchestration way, data payload limit, parallelism support, etc. Then we measure the performance of these serverless workflow services under varied experimental settings (i.e.,  different levels of activity complexity, data-flow complexity, and function complexity). Based on the results, some interesting findings, e.g, only under high data-flow complexity conditions will the performance of serverless workflow have a certain impact, can be useful. and help guide developers and cloud providers. Finally, we report a series of findings and implications to further facilitate the current practice of serverless workflow.


\balance

\bibliographystyle{ACM-Reference-Format}
\bibliography{faas_comparison}


\begin{thebibliography}{47}


\ifx \showCODEN    \undefined \def \showCODEN     #1{\unskip}     \fi
\ifx \showDOI      \undefined \def \showDOI       #1{#1}\fi
\ifx \showISBNx    \undefined \def \showISBNx     #1{\unskip}     \fi
\ifx \showISBNxiii \undefined \def \showISBNxiii  #1{\unskip}     \fi
\ifx \showISSN     \undefined \def \showISSN      #1{\unskip}     \fi
\ifx \showLCCN     \undefined \def \showLCCN      #1{\unskip}     \fi
\ifx \shownote     \undefined \def \shownote      #1{#1}          \fi
\ifx \showarticletitle \undefined \def \showarticletitle #1{#1}   \fi
\ifx \showURL      \undefined \def \showURL       {\relax}        \fi
\providecommand\bibfield[2]{#2}
\providecommand\bibinfo[2]{#2}
\providecommand\natexlab[1]{#1}
\providecommand\showeprint[2][]{arXiv:#2}

\bibitem[\protect\citeauthoryear{??}{ser}{[n.d.]}]%
        {serverlesscommunitysurvey}
 \bibinfo{year}{[n.d.]}\natexlab{}.
\newblock \bibinfo{title}{2018 serverless community survey: huge growth in
  serverless usage}.
\newblock
  \bibinfo{howpublished}{\url{https://www.serverless.com/blog/2018-serverless-community-survey-huge-growth-usage}}.
\newblock
\newblock
\shownote{Retrieved on September 10, 2020.}


\bibitem[\protect\citeauthoryear{??}{SW}{[n.d.]}]%
        {SW}
 \bibinfo{year}{[n.d.]}\natexlab{}.
\newblock \bibinfo{title}{Aliyun Serverless Workflow (in Chinese)}.
\newblock
  \bibinfo{howpublished}{\url{https://help.aliyun.com/product/113549.html?spm=a2c4g.11186623.3.1.18724f72UVzDSo}}.
\newblock
\newblock
\shownote{Retrieved on September 10, 2020.}


\bibitem[\protect\citeauthoryear{??}{Ama}{[n.d.]}]%
        {Amazon}
 \bibinfo{year}{[n.d.]}\natexlab{}.
\newblock \bibinfo{title}{Amazon}.
\newblock \bibinfo{howpublished}{\url{https://aws.amazon.com/?nc2=h_lg}}.
\newblock
\newblock
\shownote{Retrieved on September 10, 2020.}


\bibitem[\protect\citeauthoryear{??}{ASF}{[n.d.]a}]%
        {ASF}
 \bibinfo{year}{[n.d.]}\natexlab{a}.
\newblock \bibinfo{title}{AWS Step Functions documentation}.
\newblock
  \bibinfo{howpublished}{\url{https://docs.aws.amazon.com/step-functions/index.html}}.
\newblock
\newblock
\shownote{Retrieved on September 10, 2020.}


\bibitem[\protect\citeauthoryear{??}{ASF}{[n.d.]b}]%
        {ASFpayload}
 \bibinfo{year}{[n.d.]}\natexlab{b}.
\newblock \bibinfo{title}{AWS Step Functions increases payload size to 256KB}.
\newblock
  \bibinfo{howpublished}{\url{https://aws.amazon.com/about-aws/whats-new/2020/09/aws-step-functions-increases-payload-size-to-256kb/?nc1=h_ls}}.
\newblock
\newblock
\shownote{Retrieved on September 10, 2020.}


\bibitem[\protect\citeauthoryear{??}{Azu}{[n.d.]}]%
        {AzureFunctions}
 \bibinfo{year}{[n.d.]}\natexlab{}.
\newblock \bibinfo{title}{Azure Functions}.
\newblock
  \bibinfo{howpublished}{\url{https://azure.microsoft.com/en-us/services/functions/}}.
\newblock
\newblock
\shownote{Retrieved on September 10, 2020.}


\bibitem[\protect\citeauthoryear{??}{DAG}{[n.d.]}]%
        {DAGrelationship}
 \bibinfo{year}{[n.d.]}\natexlab{}.
\newblock \bibinfo{title}{The concept of DAG}.
\newblock
  \bibinfo{howpublished}{\url{https://airflow.apache.org/docs/stable/concepts.html}}.
\newblock
\newblock
\shownote{Retrieved on September 10, 2020.}


\bibitem[\protect\citeauthoryear{??}{Map}{[n.d.]}]%
        {MapReduceExample}
 \bibinfo{year}{[n.d.]}\natexlab{}.
\newblock \bibinfo{title}{ETL-DataProcessing using MapReduce}.
\newblock
  \bibinfo{howpublished}{\url{https://github.com/awesome-fnf/ETL-DataProcessing}}.
\newblock
\newblock
\shownote{Retrieved on September 10, 2020.}


\bibitem[\protect\citeauthoryear{??}{Faa}{[n.d.]}]%
        {FaaSMarket}
 \bibinfo{year}{[n.d.]}\natexlab{}.
\newblock \bibinfo{title}{Function-as-a-Service Market by User Type
  (Developer-Centric and Operator-Centric), Application (Web \& Mobile Based,
  Research \& Academic), Service Type, Deployment Model, Organization Size,
  Industry Vertical, and Region - Global Forecast to 2021}.
\newblock
  \bibinfo{howpublished}{\url{https://www.marketsandmarkets.com/Market-Reports/function-as-a-service-market-127202409.html}}.
\newblock
\newblock
\shownote{Retrieved on September 10, 2020.}


\bibitem[\protect\citeauthoryear{??}{Goo}{[n.d.]a}]%
        {Google}
 \bibinfo{year}{[n.d.]}\natexlab{a}.
\newblock \bibinfo{title}{Google}.
\newblock \bibinfo{howpublished}{\url{https://cloud.google.com/}}.
\newblock
\newblock
\shownote{Retrieved on September 10, 2020.}


\bibitem[\protect\citeauthoryear{??}{GCC}{[n.d.]}]%
        {GCC}
 \bibinfo{year}{[n.d.]}\natexlab{}.
\newblock \bibinfo{title}{Google Cloud Composer}.
\newblock
  \bibinfo{howpublished}{\url{https://cloud.google.com/composer?hl=en}}.
\newblock
\newblock
\shownote{Retrieved on September 10, 2020.}


\bibitem[\protect\citeauthoryear{??}{Mic}{[n.d.]}]%
        {Microsoft}
 \bibinfo{year}{[n.d.]}\natexlab{}.
\newblock \bibinfo{title}{Microsoft}.
\newblock \bibinfo{howpublished}{\url{https://azure.microsoft.com/en-us/}}.
\newblock
\newblock
\shownote{Retrieved on September 10, 2020.}


\bibitem[\protect\citeauthoryear{??}{SWb}{[n.d.]}]%
        {SWbestpractices}
 \bibinfo{year}{[n.d.]}\natexlab{}.
\newblock \bibinfo{title}{Serverless workflow applicable scenarios and best
  practices (in Chinese)}.
\newblock
  \bibinfo{howpublished}{\url{https://developer.aliyun.com/article/751573}}.
\newblock
\newblock
\shownote{Retrieved on September 10, 2020.}


\bibitem[\protect\citeauthoryear{??}{Sta}{[n.d.]}]%
        {StandardvsExpressWorkflows}
 \bibinfo{year}{[n.d.]}\natexlab{}.
\newblock \bibinfo{title}{Standard vs. Express Workflows}.
\newblock
  \bibinfo{howpublished}{\url{https://docs.aws.amazon.com/step-functions/latest/dg/concepts-standard-vs-express.html}}.
\newblock
\newblock
\shownote{Retrieved on September 10, 2020.}


\bibitem[\protect\citeauthoryear{??}{ADF}{[n.d.]}]%
        {ADF}
 \bibinfo{year}{[n.d.]}\natexlab{}.
\newblock \bibinfo{title}{What are Durable Functions?}
\newblock
  \bibinfo{howpublished}{\url{https://docs.microsoft.com/en-us/azure/azure-functions/durable/durable-functions-overview?tabs=csharp}}.
\newblock
\newblock
\shownote{Retrieved on September 10, 2020.}


\bibitem[\protect\citeauthoryear{??}{Goo}{[n.d.]b}]%
        {GoogleWorkflow}
 \bibinfo{year}{[n.d.]}\natexlab{b}.
\newblock \bibinfo{title}{Workflows documentation on Google Cloud}.
\newblock
  \bibinfo{howpublished}{\url{https://cloud.google.com/workflows/docs}}.
\newblock
\newblock
\shownote{Retrieved on October 07, 2020.}


\bibitem[\protect\citeauthoryear{Akkus, Chen, Rimac, Stein, Satzke, Beck,
  Aditya, and Hilt}{Akkus et~al\mbox{.}}{2018}]%
        {akkus2018sand}
\bibfield{author}{\bibinfo{person}{Istemi~Ekin Akkus},
  \bibinfo{person}{Ruichuan Chen}, \bibinfo{person}{Ivica Rimac},
  \bibinfo{person}{Manuel Stein}, \bibinfo{person}{Klaus Satzke},
  \bibinfo{person}{Andre Beck}, \bibinfo{person}{Paarijaat Aditya}, {and}
  \bibinfo{person}{Volker Hilt}.} \bibinfo{year}{2018}\natexlab{}.
\newblock \showarticletitle{SAND: towards high-performance serverless
  computing}. In \bibinfo{booktitle}{\emph{Proceedings of the 2018 USENIX
  Annual Technical Conference}}. \bibinfo{pages}{923--935}.
\newblock


\bibitem[\protect\citeauthoryear{Bila, Dettori, Kanso, Watanabe, and
  Youssef}{Bila et~al\mbox{.}}{2017}]%
        {bila2017leveraging}
\bibfield{author}{\bibinfo{person}{Nilton Bila}, \bibinfo{person}{Paolo
  Dettori}, \bibinfo{person}{Ali Kanso}, \bibinfo{person}{Yuji Watanabe}, {and}
  \bibinfo{person}{Alaa Youssef}.} \bibinfo{year}{2017}\natexlab{}.
\newblock \showarticletitle{Leveraging the serverless architecture for securing
  linux containers}. In \bibinfo{booktitle}{\emph{Proceedings of 37th IEEE
  International Conference on Distributed Computing Systems Workshops}}.
  \bibinfo{pages}{401--404}.
\newblock


\bibitem[\protect\citeauthoryear{Cardoso}{Cardoso}{2006}]%
        {cardoso2006approaches}
\bibfield{author}{\bibinfo{person}{Jorge Cardoso}.}
  \bibinfo{year}{2006}\natexlab{}.
\newblock \showarticletitle{Approaches to compute workflow complexity}. In
  \bibinfo{booktitle}{\emph{Proceedings of Role of Business Processes in
  Service Oriented Architectures (Dagstuhl Seminar Proceedings)}}. Schloss
  Dagstuhl-Leibniz-Zentrum f{\"u}r Informatik.
\newblock


\bibitem[\protect\citeauthoryear{Carreira, Fonseca, Tumanov, Zhang, and
  Katz}{Carreira et~al\mbox{.}}{2018}]%
        {carreira2018case}
\bibfield{author}{\bibinfo{person}{Joao Carreira}, \bibinfo{person}{Pedro
  Fonseca}, \bibinfo{person}{Alexey Tumanov}, \bibinfo{person}{Andrew Zhang},
  {and} \bibinfo{person}{Randy Katz}.} \bibinfo{year}{2018}\natexlab{}.
\newblock \showarticletitle{A case for serverless machine learning}. In
  \bibinfo{booktitle}{\emph{Proceedings of Workshop on Systems for ML and Open
  Source Software at NeurIPS}}, Vol.~\bibinfo{volume}{2018}.
\newblock


\bibitem[\protect\citeauthoryear{Chard, Chard, Alt, Parkinson, Tuecke, and
  Foster}{Chard et~al\mbox{.}}{2017}]%
        {chard2017ripple}
\bibfield{author}{\bibinfo{person}{Ryan Chard}, \bibinfo{person}{Kyle Chard},
  \bibinfo{person}{Jason Alt}, \bibinfo{person}{Dilworth~Y Parkinson},
  \bibinfo{person}{Steve Tuecke}, {and} \bibinfo{person}{Ian Foster}.}
  \bibinfo{year}{2017}\natexlab{}.
\newblock \showarticletitle{Ripple: home automation for research data
  management}. In \bibinfo{booktitle}{\emph{Proceedings of 37th IEEE
  International Conference on Distributed Computing Systems Workshops}}.
  \bibinfo{pages}{389--394}.
\newblock


\bibitem[\protect\citeauthoryear{Chen, Cao, Liu, Wang, Wang, and Liu}{Chen
  et~al\mbox{.}}{2020}]%
        {add14}
\bibfield{author}{\bibinfo{person}{Zhenpeng Chen}, \bibinfo{person}{Yanbin
  Cao}, \bibinfo{person}{Yuanqiang Liu}, \bibinfo{person}{Haoyu Wang},
  \bibinfo{person}{Tao Wang}, {and} \bibinfo{person}{Xuanzhe Liu}.}
  \bibinfo{year}{2020}\natexlab{}.
\newblock \showarticletitle{A comprehensive study on challenges in deploying
  deep learning based software}. In \bibinfo{booktitle}{\emph{Proceedings of
  the 28th ACM Joint Meeting on European Software Engineering Conference and
  Symposium on the Foundations of Software Engineering}}.
  \bibinfo{pages}{750--762}.
\newblock


\bibitem[\protect\citeauthoryear{Datta, Kumar, Morris, Grace, Rahmati, and
  Bates}{Datta et~al\mbox{.}}{2020}]%
        {datta2020valve}
\bibfield{author}{\bibinfo{person}{Pubali Datta}, \bibinfo{person}{Prabuddha
  Kumar}, \bibinfo{person}{Tristan Morris}, \bibinfo{person}{Michael Grace},
  \bibinfo{person}{Amir Rahmati}, {and} \bibinfo{person}{Adam Bates}.}
  \bibinfo{year}{2020}\natexlab{}.
\newblock \showarticletitle{Valve: Securing Function Workflows on Serverless
  Computing Platforms}. In \bibinfo{booktitle}{\emph{Proceedings of the 29th
  International Conference on World Wide Web}}. \bibinfo{pages}{939--950}.
\newblock


\bibitem[\protect\citeauthoryear{de~Lara, Gomes, Langridge, Mortazavi, and
  Roodi}{de~Lara et~al\mbox{.}}{2016}]%
        {de2016hierarchical}
\bibfield{author}{\bibinfo{person}{Eyal de Lara}, \bibinfo{person}{Carolina~S
  Gomes}, \bibinfo{person}{Steve Langridge}, \bibinfo{person}{S~Hossein
  Mortazavi}, {and} \bibinfo{person}{Meysam Roodi}.}
  \bibinfo{year}{2016}\natexlab{}.
\newblock \showarticletitle{Hierarchical serverless computing for the mobile
  edge}. In \bibinfo{booktitle}{\emph{Proceedings of IEEE/ACM Symposium on Edge
  Computing}}. \bibinfo{pages}{109--110}.
\newblock


\bibitem[\protect\citeauthoryear{Fouladi, Wahby, Shacklett, Balasubramaniam,
  Zeng, Bhalerao, Sivaraman, Porter, and Winstein}{Fouladi
  et~al\mbox{.}}{2017}]%
        {fouladi2017encoding}
\bibfield{author}{\bibinfo{person}{Sadjad Fouladi}, \bibinfo{person}{Riad~S
  Wahby}, \bibinfo{person}{Brennan Shacklett},
  \bibinfo{person}{Karthikeyan~Vasuki Balasubramaniam},
  \bibinfo{person}{William Zeng}, \bibinfo{person}{Rahul Bhalerao},
  \bibinfo{person}{Anirudh Sivaraman}, \bibinfo{person}{George Porter}, {and}
  \bibinfo{person}{Keith Winstein}.} \bibinfo{year}{2017}\natexlab{}.
\newblock \showarticletitle{Encoding, fast and slow: low-latency video
  processing using thousands of tiny threads}. In
  \bibinfo{booktitle}{\emph{Proceedings of 14th USENIX Symposium on Networked
  Systems Design and Implementation}}. \bibinfo{pages}{363--376}.
\newblock


\bibitem[\protect\citeauthoryear{Huang, Liu, Ma, Lu, Zhang, and Xiong}{Huang
  et~al\mbox{.}}{2016}]%
        {add3}
\bibfield{author}{\bibinfo{person}{Gang Huang}, \bibinfo{person}{Xuanzhe Liu},
  \bibinfo{person}{Yun Ma}, \bibinfo{person}{Xuan Lu}, \bibinfo{person}{Ying
  Zhang}, {and} \bibinfo{person}{Yingfei Xiong}.}
  \bibinfo{year}{2016}\natexlab{}.
\newblock \showarticletitle{Programming situational mobile web applications
  with cloud-mobile convergence: An internetware-oriented approach}.
\newblock \bibinfo{journal}{\emph{IEEE Transactions on Services Computing}}
  \bibinfo{volume}{12}, \bibinfo{number}{1} (\bibinfo{year}{2016}),
  \bibinfo{pages}{6--19}.
\newblock


\bibitem[\protect\citeauthoryear{Huang, Luo, Wu, Ma, Zhang, and Liu}{Huang
  et~al\mbox{.}}{2019}]%
        {add7}
\bibfield{author}{\bibinfo{person}{Gang Huang}, \bibinfo{person}{Chaoran Luo},
  \bibinfo{person}{Kaidong Wu}, \bibinfo{person}{Yun Ma}, \bibinfo{person}{Ying
  Zhang}, {and} \bibinfo{person}{Xuanze Liu}.} \bibinfo{year}{2019}\natexlab{}.
\newblock \showarticletitle{Software-defined infrastructure for decentralized
  data lifecycle governance: Principled design and open challenges}. In
  \bibinfo{booktitle}{\emph{2019 IEEE 39th International Conference on
  Distributed Computing Systems (ICDCS)}}. IEEE, \bibinfo{pages}{1674--1683}.
\newblock


\bibitem[\protect\citeauthoryear{Huang, Xu, Lin, Liu, Ma, Pushp, and Liu}{Huang
  et~al\mbox{.}}{2017}]%
        {add11}
\bibfield{author}{\bibinfo{person}{Gang Huang}, \bibinfo{person}{Mengwei Xu},
  \bibinfo{person}{Felix~Xiaozhu Lin}, \bibinfo{person}{Yunxin Liu},
  \bibinfo{person}{Yun Ma}, \bibinfo{person}{Saumay Pushp}, {and}
  \bibinfo{person}{Xuanzhe Liu}.} \bibinfo{year}{2017}\natexlab{}.
\newblock \showarticletitle{Shuffledog: Characterizing and adapting
  user-perceived latency of android apps}.
\newblock \bibinfo{journal}{\emph{IEEE Transactions on Mobile Computing}}
  \bibinfo{volume}{16}, \bibinfo{number}{10} (\bibinfo{year}{2017}),
  \bibinfo{pages}{2913--2926}.
\newblock


\bibitem[\protect\citeauthoryear{Jonas, Pu, Venkataraman, Stoica, and
  Recht}{Jonas et~al\mbox{.}}{2017}]%
        {jonas2017occupy}
\bibfield{author}{\bibinfo{person}{Eric Jonas}, \bibinfo{person}{Qifan Pu},
  \bibinfo{person}{Shivaram Venkataraman}, \bibinfo{person}{Ion Stoica}, {and}
  \bibinfo{person}{Benjamin Recht}.} \bibinfo{year}{2017}\natexlab{}.
\newblock \showarticletitle{Occupy the cloud: distributed computing for the
  99\%}. In \bibinfo{booktitle}{\emph{Proceedings of 2017 Symposium on Cloud
  Computing}}. \bibinfo{pages}{445--451}.
\newblock


\bibitem[\protect\citeauthoryear{Jonas, Schleier-Smith, Sreekanti, Tsai,
  Khandelwal, Pu, Shankar, Carreira, Krauth, Yadwadkar, Gonzalez, Popa, Stoica,
  and Patterson}{Jonas et~al\mbox{.}}{2019}]%
        {jonas2019cloud}
\bibfield{author}{\bibinfo{person}{Eric Jonas}, \bibinfo{person}{Johann
  Schleier-Smith}, \bibinfo{person}{Vikram Sreekanti},
  \bibinfo{person}{Chia-Che Tsai}, \bibinfo{person}{Anurag Khandelwal},
  \bibinfo{person}{Qifan Pu}, \bibinfo{person}{Vaishaal Shankar},
  \bibinfo{person}{Joao Carreira}, \bibinfo{person}{Karl Krauth},
  \bibinfo{person}{Neeraja Yadwadkar}, \bibinfo{person}{Joseph~E. Gonzalez},
  \bibinfo{person}{Raluca~Ada Popa}, \bibinfo{person}{Ion Stoica}, {and}
  \bibinfo{person}{David~A. Patterson}.} \bibinfo{year}{2019}\natexlab{}.
\newblock \showarticletitle{Cloud programming simplified: A berkeley view on
  serverless computing}.
\newblock \bibinfo{journal}{\emph{arXiv preprint arXiv:1902.03383}}
  (\bibinfo{year}{2019}).
\newblock


\bibitem[\protect\citeauthoryear{Liu, Ma, Dong, Liu, Xie, and Huang}{Liu
  et~al\mbox{.}}{2016a}]%
        {add10}
\bibfield{author}{\bibinfo{person}{Xuanzhe Liu}, \bibinfo{person}{Yun Ma},
  \bibinfo{person}{Shuailiang Dong}, \bibinfo{person}{Yunxin Liu},
  \bibinfo{person}{Tao Xie}, {and} \bibinfo{person}{Gang Huang}.}
  \bibinfo{year}{2016}\natexlab{a}.
\newblock \showarticletitle{ReWAP: Reducing redundant transfers for mobile web
  browsing via app-specific resource packaging}.
\newblock \bibinfo{journal}{\emph{IEEE Transactions on Mobile Computing}}
  \bibinfo{volume}{16}, \bibinfo{number}{9} (\bibinfo{year}{2016}),
  \bibinfo{pages}{2625--2638}.
\newblock


\bibitem[\protect\citeauthoryear{Liu, Ma, Huang, Zhao, Mei, and Liu}{Liu
  et~al\mbox{.}}{2014}]%
        {add4}
\bibfield{author}{\bibinfo{person}{Xuanzhe Liu}, \bibinfo{person}{Yun Ma},
  \bibinfo{person}{Gang Huang}, \bibinfo{person}{Junfeng Zhao},
  \bibinfo{person}{Hong Mei}, {and} \bibinfo{person}{Yunxin Liu}.}
  \bibinfo{year}{2014}\natexlab{}.
\newblock \showarticletitle{Data-driven composition for service-oriented
  situational web applications}.
\newblock \bibinfo{journal}{\emph{IEEE Transactions on Services Computing}}
  \bibinfo{volume}{8}, \bibinfo{number}{1} (\bibinfo{year}{2014}),
  \bibinfo{pages}{2--16}.
\newblock


\bibitem[\protect\citeauthoryear{Liu, Ma, and Lin}{Liu et~al\mbox{.}}{2018a}]%
        {add2}
\bibfield{author}{\bibinfo{person}{Xuanzhe Liu}, \bibinfo{person}{Yun Ma},
  {and} \bibinfo{person}{Felix~Xiaozhu Lin}.} \bibinfo{year}{2018}\natexlab{a}.
\newblock \showarticletitle{Rethinking Resource Management in Mobile Web:
  Measurement, Deployment, and Runtime}. In \bibinfo{booktitle}{\emph{2018 IEEE
  38th International Conference on Distributed Computing Systems (ICDCS)}}.
  IEEE, \bibinfo{pages}{1347--1356}.
\newblock


\bibitem[\protect\citeauthoryear{Liu, Ma, Wang, Liu, Xie, and Huang}{Liu
  et~al\mbox{.}}{2016b}]%
        {add9}
\bibfield{author}{\bibinfo{person}{Xuanzhe Liu}, \bibinfo{person}{Yun Ma},
  \bibinfo{person}{Xinyang Wang}, \bibinfo{person}{Yunxin Liu},
  \bibinfo{person}{Tao Xie}, {and} \bibinfo{person}{Gang Huang}.}
  \bibinfo{year}{2016}\natexlab{b}.
\newblock \showarticletitle{SWAROVsky: Optimizing resource loading for mobile
  web browsing}.
\newblock \bibinfo{journal}{\emph{IEEE Transactions on Mobile Computing}}
  \bibinfo{volume}{16}, \bibinfo{number}{10} (\bibinfo{year}{2016}),
  \bibinfo{pages}{2941--2954}.
\newblock


\bibitem[\protect\citeauthoryear{Liu, Sun, and Huang}{Liu
  et~al\mbox{.}}{2019}]%
        {add6}
\bibfield{author}{\bibinfo{person}{Xuanzhe Liu}, \bibinfo{person}{Sam~Xun Sun},
  {and} \bibinfo{person}{Gang Huang}.} \bibinfo{year}{2019}\natexlab{}.
\newblock \showarticletitle{Decentralized Services Computing Paradigm for
  Blockchain-Based Data Governance: Programmability, Interoperability, and
  Intelligence}.
\newblock \bibinfo{journal}{\emph{IEEE Transactions on Services Computing}}
  \bibinfo{volume}{13}, \bibinfo{number}{2} (\bibinfo{year}{2019}),
  \bibinfo{pages}{343--355}.
\newblock


\bibitem[\protect\citeauthoryear{Liu, Yu, Ma, Huang, Mei, and Liu}{Liu
  et~al\mbox{.}}{2018b}]%
        {add8}
\bibfield{author}{\bibinfo{person}{Xuanzhe Liu}, \bibinfo{person}{Meihua Yu},
  \bibinfo{person}{Yun Ma}, \bibinfo{person}{Gang Huang}, \bibinfo{person}{Hong
  Mei}, {and} \bibinfo{person}{Yunxin Liu}.} \bibinfo{year}{2018}\natexlab{b}.
\newblock \showarticletitle{i-Jacob: An internetware-oriented approach to
  optimizing computation-intensive mobile web browsing}.
\newblock \bibinfo{journal}{\emph{ACM Transactions on Internet Technology}}
  \bibinfo{volume}{18}, \bibinfo{number}{2} (\bibinfo{year}{2018}),
  \bibinfo{pages}{1--23}.
\newblock


\bibitem[\protect\citeauthoryear{L{\'o}pez, Arjona, Samp{\'e}, Slominski, and
  Villard}{L{\'o}pez et~al\mbox{.}}{2020}]%
        {lopez2020triggerflow}
\bibfield{author}{\bibinfo{person}{Pedro~Garc{\'\i}a L{\'o}pez},
  \bibinfo{person}{Aitor Arjona}, \bibinfo{person}{Josep Samp{\'e}},
  \bibinfo{person}{Aleksander Slominski}, {and} \bibinfo{person}{Lionel
  Villard}.} \bibinfo{year}{2020}\natexlab{}.
\newblock \showarticletitle{Triggerflow: trigger-based orchestration of
  serverless workflows}. In \bibinfo{booktitle}{\emph{Proceedings of the 14th
  ACM International Conference on Distributed and Event-based Systems}}.
  \bibinfo{pages}{3--14}.
\newblock


\bibitem[\protect\citeauthoryear{L{\'o}pez, S{\'a}nchez-Artigas, Par{\'\i}s,
  Pons, Ollobarren, and Pinto}{L{\'o}pez et~al\mbox{.}}{2018}]%
        {lopez2018comparison}
\bibfield{author}{\bibinfo{person}{Pedro~Garc{\'\i}a L{\'o}pez},
  \bibinfo{person}{Marc S{\'a}nchez-Artigas}, \bibinfo{person}{Gerard
  Par{\'\i}s}, \bibinfo{person}{Daniel~Barcelona Pons},
  \bibinfo{person}{{\'A}lvaro~Ruiz Ollobarren}, {and}
  \bibinfo{person}{David~Arroyo Pinto}.} \bibinfo{year}{2018}\natexlab{}.
\newblock \showarticletitle{Comparison of faas orchestration systems}. In
  \bibinfo{booktitle}{\emph{Proceedings of 2018 IEEE/ACM International
  Conference on Utility and Cloud Computing Companion}}.
  \bibinfo{pages}{148--153}.
\newblock


\bibitem[\protect\citeauthoryear{Ma, Liu, Zhang, Xiang, Liu, and Xie}{Ma
  et~al\mbox{.}}{2015}]%
        {add13}
\bibfield{author}{\bibinfo{person}{Yun Ma}, \bibinfo{person}{Xuanzhe Liu},
  \bibinfo{person}{Shuhui Zhang}, \bibinfo{person}{Ruirui Xiang},
  \bibinfo{person}{Yunxin Liu}, {and} \bibinfo{person}{Tao Xie}.}
  \bibinfo{year}{2015}\natexlab{}.
\newblock \showarticletitle{Measurement and analysis of mobile web cache
  performance}. In \bibinfo{booktitle}{\emph{Proceedings of the 24th
  International Conference on World Wide Web (WWW)}}.
  \bibinfo{pages}{691--701}.
\newblock


\bibitem[\protect\citeauthoryear{Malawski, Gajek, Zima, Balis, and
  Figiela}{Malawski et~al\mbox{.}}{2020}]%
        {Malawski2020}
\bibfield{author}{\bibinfo{person}{Maciej Malawski}, \bibinfo{person}{Adam
  Gajek}, \bibinfo{person}{Adam Zima}, \bibinfo{person}{Bartosz Balis}, {and}
  \bibinfo{person}{Kamil Figiela}.} \bibinfo{year}{2020}\natexlab{}.
\newblock \showarticletitle{Serverless execution of scientific workflows:
  experiments with {HyperFlow}, {AWS} Lambda and Google Cloud Functions}.
\newblock \bibinfo{journal}{\emph{Future Generation Computer Systems}}
  \bibinfo{volume}{110} (\bibinfo{year}{2020}), \bibinfo{pages}{502--514}.
\newblock


\bibitem[\protect\citeauthoryear{McGrath and Brenner}{McGrath and
  Brenner}{2017}]%
        {mcgrath2017serverless}
\bibfield{author}{\bibinfo{person}{Garrett McGrath} {and}
  \bibinfo{person}{Paul~R Brenner}.} \bibinfo{year}{2017}\natexlab{}.
\newblock \showarticletitle{Serverless computing: design, implementation, and
  performance}. In \bibinfo{booktitle}{\emph{Proceedings of 2017 IEEE 37th
  International Conference on Distributed Computing Systems Workshops}}.
  \bibinfo{pages}{405--410}.
\newblock


\bibitem[\protect\citeauthoryear{Shahrad, Fonseca, Goiri, Chaudhry, Batum,
  Cooke, Laureano, Tresness, Russinovich, and Bianchini}{Shahrad
  et~al\mbox{.}}{2020}]%
        {shahradserverless2020}
\bibfield{author}{\bibinfo{person}{Mohammad Shahrad}, \bibinfo{person}{Rodrigo
  Fonseca}, \bibinfo{person}{{\'I}{\~n}igo Goiri}, \bibinfo{person}{Gohar
  Chaudhry}, \bibinfo{person}{Paul Batum}, \bibinfo{person}{Jason Cooke},
  \bibinfo{person}{Eduardo Laureano}, \bibinfo{person}{Colby Tresness},
  \bibinfo{person}{Mark Russinovich}, {and} \bibinfo{person}{Ricardo
  Bianchini}.} \bibinfo{year}{2020}\natexlab{}.
\newblock \showarticletitle{Serverless in the wild: characterizing and
  optimizing the serverless workload at a large cloud provider}. In
  \bibinfo{booktitle}{\emph{Proceedings of the 2020 USENIX Annual Technical
  Conference}}. \bibinfo{pages}{205--218}.
\newblock


\bibitem[\protect\citeauthoryear{Wang, Li, Zhang, Ristenpart, and Swift}{Wang
  et~al\mbox{.}}{2018}]%
        {wang2018peeking}
\bibfield{author}{\bibinfo{person}{Liang Wang}, \bibinfo{person}{Mengyuan Li},
  \bibinfo{person}{Yinqian Zhang}, \bibinfo{person}{Thomas Ristenpart}, {and}
  \bibinfo{person}{Michael Swift}.} \bibinfo{year}{2018}\natexlab{}.
\newblock \showarticletitle{Peeking behind the curtains of serverless
  platforms}. In \bibinfo{booktitle}{\emph{Proceedings of the 2018 USENIX
  Annual Technical Conference}}. \bibinfo{pages}{133--146}.
\newblock


\bibitem[\protect\citeauthoryear{Xu, Jiang, Luo, Sun, An, Huang, and Liu}{Xu
  et~al\mbox{.}}{2020a}]%
        {add1}
\bibfield{author}{\bibinfo{person}{Chenren Xu}, \bibinfo{person}{Shuang Jiang},
  \bibinfo{person}{Guojie Luo}, \bibinfo{person}{Guangyu Sun},
  \bibinfo{person}{Ning An}, \bibinfo{person}{Gang Huang}, {and}
  \bibinfo{person}{Xuanzhe Liu}.} \bibinfo{year}{2020}\natexlab{a}.
\newblock \showarticletitle{The Case for FPGA-based Edge Computing}.
\newblock \bibinfo{journal}{\emph{IEEE Transactions on Mobile Computing}}
  (\bibinfo{year}{2020}).
\newblock


\bibitem[\protect\citeauthoryear{Xu, Zhang, Liu, Huang, Liu, and Lin}{Xu
  et~al\mbox{.}}{2020b}]%
        {add5}
\bibfield{author}{\bibinfo{person}{Mengwei Xu}, \bibinfo{person}{Xiwen Zhang},
  \bibinfo{person}{Yunxin Liu}, \bibinfo{person}{Gang Huang},
  \bibinfo{person}{Xuanzhe Liu}, {and} \bibinfo{person}{Felix~Xiaozhu Lin}.}
  \bibinfo{year}{2020}\natexlab{b}.
\newblock \showarticletitle{Approximate query service on autonomous iot
  cameras}. In \bibinfo{booktitle}{\emph{Proceedings of the 18th International
  Conference on Mobile Systems, Applications, and Services (MobiSys)}}.
  \bibinfo{pages}{191--205}.
\newblock


\bibitem[\protect\citeauthoryear{Yuan and Yang}{Yuan and Yang}{2019}]%
        {yuankmeansresearch2019}
\bibfield{author}{\bibinfo{person}{Chunhui Yuan} {and} \bibinfo{person}{Haitao
  Yang}.} \bibinfo{year}{2019}\natexlab{}.
\newblock \showarticletitle{Research on K-value selection method of K-means
  clustering algorithm}.
\newblock \bibinfo{journal}{\emph{J—Multidisciplinary Scientific Journal}}
  \bibinfo{volume}{2}, \bibinfo{number}{2} (\bibinfo{year}{2019}),
  \bibinfo{pages}{226--235}.
\newblock


\bibitem[\protect\citeauthoryear{Zhang, Liu, Liu, and Li}{Zhang
  et~al\mbox{.}}{2017}]%
        {add12}
\bibfield{author}{\bibinfo{person}{Yifan Zhang}, \bibinfo{person}{Yunxin Liu},
  \bibinfo{person}{Xuanzhe Liu}, {and} \bibinfo{person}{Qun Li}.}
  \bibinfo{year}{2017}\natexlab{}.
\newblock \showarticletitle{Enabling accurate and efficient modeling-based CPU
  power estimation for smartphones}. In \bibinfo{booktitle}{\emph{2017 IEEE/ACM
  25th International Symposium on Quality of Service (IWQoS)}}. IEEE,
  \bibinfo{pages}{1--10}.
\newblock


\end{thebibliography}

\end{document}